\documentclass[superscriptaddress,preprint,preprintnumbers,amsmath,amssymb]{revtex4-1}

\usepackage{graphicx,dcolumn,bm,color,xcolor,braket,tikz,setspace}
\usepackage{soul} 
\usepackage[caption=false]{subfig}
\def\kt{{k_{\rm B}T}}
\def\k{\kappa}
\def\a{\alpha}

\def\DD{\Delta}
\def\w{\omega}

\def\bk{{\bf k}}
\def\bK{{\bf K}}
\def\bG{{\bf G}}
\def\bq{{\bf q} }

\def\bR{{\bf R} }

\def\ve{\varepsilon}

\def\<{\langle}
\def\>{\rangle}
\def\D{\partial}
\def\Np{{N_{\rm uc}}}
\let\hide\iffalse

\newcommand{\br}{\mathbf r}

\newcommand{\aibte}{{\textit{ai}BTE}}

\begin{document}

\setstretch{1.2}

\title{Electron-phonon physics from first principles using the EPW code}

\def\utoden{Oden Institute for Computational Engineering and Sciences, The University of Texas at Austin, Austin, Texas 78712, USA}
\def\utphysics{Department of Physics, The University of Texas at Austin, Austin, Texas 78712, USA}
\def\umacau{Institute of Applied Physics and Materials Engineering, University of Macau, Macao SAR 999078, P. R. China}
\def\umich{Department of Materials Science and Engineering, University of Michigan, Ann Arbor, Michigan, 48109, USA}
\def\urennes{University of Rennes, INSA Rennes, CNRS, Institut FOTON-UMR 6082, F-35000 Rennes, France}
\def\uclouvain{Institute of Condensed Matter and Nanosciences, Universit\'e Catholique de Louvain, BE-1348 Louvain-la-Neuve, Belgium}
\def\kcl{Department of Physics, King's College London, Strand, London WC2R 2LS, United Kingdom}
\def\iit{Istituto Italiano di Tecnologia, Graphene Labs, Via Morego 30, I-16163 Genova, Italy}
\def\buni{Department of Physics, Applied Physics and Astronomy, Binghamton University-SUNY, Binghamton, NY 13902, USA}
\def\unib{Department of Chemistry, University at Buffalo, Buffalo, New York 14260, USA}

\author{Hyungjun Lee}\email{hyungjun.lee@austin.utexas.edu}
  \affiliation{\utoden}
  \affiliation{\utphysics}
\author{Samuel Ponc\'e}
  \affiliation{\uclouvain}
\author{Kyle Bushick}
  \affiliation{\umich}
\author{Samad Hajinazar} 
  \altaffiliation[Current address: ]{\unib} 
  \affiliation{\buni}
\author{Jon Lafuente-Bartolome}
  \affiliation{\utoden}
  \affiliation{\utphysics}
\author{Joshua Leveillee}
  \affiliation{\utoden}
  \affiliation{\utphysics}
\author{Chao Lian}
  \affiliation{\utoden}
  \affiliation{\utphysics}
\author{Francesco Macheda}
  \affiliation{\kcl}
  \affiliation{\iit}
\author{Hari Paudyal}
  \affiliation{\buni}
\author{Weng Hong Sio}
  \affiliation{\umacau}
  \affiliation{\utoden}
\author{Marios Zacharias}
  \affiliation{\urennes}
\author{Xiao Zhang}
  \affiliation{\umich}
\author{Nicola Bonini}
  \affiliation{\kcl}
\author{Emmanouil Kioupakis}
  \affiliation{\umich}
\author{Elena R. Margine}
  \affiliation{\buni}
\author{Feliciano Giustino}\email{fgiustino@oden.utexas.edu}
  \affiliation{\utoden}
  \affiliation{\utphysics}

\date{\today}


\maketitle
\begin{center}
Abstract
\end{center}
EPW is an open-source software for \textit{ab initio} calculations of electron-phonon interactions and related materials properties. The code combines density functional perturbation theory and maximally-localized Wannier functions to efficiently compute electron-phonon coupling matrix elements on ultra-fine Brillouin zone grids. This data is employed for predictive calculations of temperature-dependent properties and phonon-assisted quantum processes in bulk solids and low-dimensional materials. Here, we report on significant new developments in the code that occurred during the period 2016-2022, namely: a transport module for the calculation of charge carrier mobility and conductivity under electric and magnetic fields within the \textit{ab initio} Boltzmann transport equation; a superconductivity module for the calculation of critical temperature and gap structure in phonon-mediated superconductors within the \textit{ab initio} anisotropic multi-band Eliashberg theory; an optics module for calculations of phonon-assisted indirect transitions; a module for the calculation of small and large polarons without supercells using the \textit{ab initio} polaron equations; and a module for calculating electron-phonon couplings, band structure renormalization, and temperature-dependent optical spectra using the special displacement method. For each capability, we outline the methodology and implementation, and provide example calculations. We describe recent code refactoring to prepare EPW for exascale architectures, we discuss efficient parallelization strategies, and report on extreme parallel scaling tests.


\newpage

\section{introduction}

The coupling between electrons and phonons is one of the most intensively studied fermion-boson interactions in condensed matter physics. It is responsible for a number of physical phenomena in solids such as conventional superconductivity, temperature-dependent resistivity in metals and mobility in semiconductors, the formation of polarons, and phonon-assisted optical processes, to name a few~\cite{Giustino2017z, Ziman1960}. Predictive non-empirical calculations of electron-phonon interactions play an essential role in elucidating a variety of materials properties and their temperature dependence.

During the past three decades, first-principles calculations of electron-phonon interactions have been made possible by the development of density functional theory (DFT)~\cite{Hohenberg1964, Kohn1965} and density-functional perturbation theory (DFPT)~\cite{Baroni1987,Savrasov1992,Gonze1997}. However, the computational cost of these calculations is high, as they involve the evaluation of Brillouin zone integrals which typically require a very fine sampling of the crystal momenta of electrons and phonons. For instance, in DFPT calculations, every single phonon wavevector and vibrational mode requires the solution of Sternheimer-type equations that carry a computational complexity comparable to a DFT total energy calculation~\cite{Baroni2001s}. As a result, computing properties relating to the electron-phonon interaction usually entails a computational workload that is orders of magnitude more costly than standard DFT calculations, thus rendering direct \textit{ab initio} calculations of these quantities computationally challenging.

Several methods have been proposed to address this challenge~\cite{Giustino2017z}. Among those, one of the authors proposed to combine DFPT with maximally-localized Wannier functions (MLWFs)~\cite{Marzari2012o} to perform a physics-based interpolation via a generalized Fourier transformation~\cite{Giustino2007l}. This method exploits the spatial localization of the electron-phonon matrix elements in the Wannier representation, and enables efficient calculations of electron-phonon matrix elements on ultra-dense momentum grids while retaining the accuracy of DFPT.

The \texttt{EPW} code builds upon this methodology to compute a number of properties relating to electron-phonon interactions and temperature-dependent materials properties. \texttt{EPW} is the first open-source \textit{ab initio} software devoted to electron-phonon interactions, and has been actively developed for over 16 years. It was publicly released in 2010 under a GNU General Public License (GPL) \cite{GPL}, and is being distributed within the \texttt{Quantum ESPRESSO} materials simulation suite \cite{giannozzi2017o} since 2016. In addition to \texttt{EPW}, several software projects have been developed in recent years to address electron-phonon interactions, including: \texttt{Perturbo} \cite{Zhou2021}, \texttt{elphbolt} \cite{Protik2022}, \texttt{Phoebe} \cite{Cepellotti2022}, and \texttt{EPIq} \cite{EPIq}. All these packages rely on the same interpolation methodology employed in \texttt{EPW} \cite{Giustino2007l}.

Two prior manuscripts \cite{Noffsinger2010,Ponce2016a} describe the status of the \texttt{EPW} code until 2016. This manuscript aims to describe progress on the code that has occurred since 2016 as well as ongoing efforts, leading to the \texttt{EPW v6} release. New functionalities involve: a magneto-transport module for the calculation of the conductivity in metals as well as the drift and Hall mobility in semiconductors; a module for the calculation of small and large polarons without using supercells; a module for the solution of the full-band Eliashberg equations for superconductors; a module for calculations of phonon-assisted indirect optical processes; and a module for calculations of finite-temperature electronic and optical properties via the special displacement method. The code has been refactored to keep up with contemporary developments in high-performance computing (HPC) architectures, for example via the introduction of hybrid two-level MPI (Message Passing Interface,~\cite{mpi40}) and OpenMP (Open Multi-Processing,~\cite{dagum1998openmp}) parallelization, and the addition of parallel I/O via parallel HDF5 (Hierarchical Data Format 5,~\cite{hdf5}).

The manuscript is organized as follows. In Sec.~\ref{sec.methodology} we set up the notation employed throughout this manuscript, we review the conceptual basis of electron-phonon interpolation using DFPT and MLWFs, and we discuss the extension of this methodology to the case of polar materials with long-range Coulomb interactions. Sec.~\ref{sec.functionalities} describes the new or expanded functionalities available in the latest release of the \texttt{EPW} code. In particular, in Sec.~\ref{sec.transport} we discuss calculations of carrier transport within the \textit{ab initio} Boltzmann transport equation. We consider both electric and magnetic fields, as well as phonon-limited and charged defect-limited transport. Section~\ref{sec.sc} is devoted to calculations of the superconducting critical temperature and superconducting gap function. We discuss elementary calculations based on the semi-empirical McMillan equation, the isotropic Eliashberg theory, and the fully-anisotropic full-band Eliashberg theory. In Sec.~\ref{sec.polarons} we outline the methodology for computing small and large polarons from DFPT without resorting to large supercells. Section~\ref{sec.indabs} covers the formalism for the calculation of optical absorption spectra including phonon-assisted indirect transitions. Section~\ref{sec.sdm} is devoted to the special displacement method, which offers an alternative and complementary strategy to computing electron-phonon properties without employing Wannier-Fourier interpolation. In Sec.~\ref{sec.hpc} we discuss recent advances in the computational algorithms, parallelization, and I/O, and we report on extreme parallel scaling benchmarks. In Sec.~\ref{sec.futuredirections} we discuss possible future directions for the \texttt{EPW} software project, and in Sec.~\ref{sec.conclusions} we draw our conclusions. 

\section{Methodology}\label{sec.methodology}

The key element of any calculation of electron-phonon interactions and related materials properties is the electron-phonon matrix element. The \texttt{EPW} code employs physics-based Wannier-Fourier interpolation to compute electron-phonon matrix elements accurately and efficiently, and uses these matrix elements as the starting point for advanced many-body calculations of electronic, optical, and transport properties of solids including temperature dependence and phonon-assisted quantum processes.

In the following two sections we introduce the notations and conventions employed throughout the manuscript (Sec.~\ref{basicdef}), we provide a brief overview of the general principles of Wannier-Fourier interpolation of electron-phonon matrix elements (Sec.~\ref{sec.epwtheory}), and we outline the extension of this interpolation scheme to the case of polar materials (Sec.~\ref{sec.longrange}). A more comprehensive description of this methodology including detailed derivations can be found in Ref.~\cite{Giustino2007l}.

\subsection{Notation and definitions}
\label{basicdef}

In the study of the electronic structure and lattice dynamics of crystalline solids, infinitely-extended crystals are described by considering a periodic Born-von K\'arm\'an (BvK) supercell. This supercell consists of
$\Np=N_1\!\times \!N_2\! \times\! N_3$ primitive unit cells of the crystal, and is subject to periodic boundary conditions. The $p$-th unit cell within the BvK supercell is identified by the direct lattice vector ${\bf R}_p$, with $p=1,\dots,\Np$. The dual lattice of the direct lattice vectors $\bR_p$ in reciprocal space defines a uniform grid of $\Np=N_1\!\times \!N_2\! \times\! N_3$ wavevectors in the Brillouin zone. Throughout the manuscript, we use $\bk$ and $\bq$ to denote electron and phonon wavevectors in the crystal Brillouin zone, respectively, and $\bK$ to denote electron wavevectors in the Brillouin zone of the BvK supercell.

We use $\psi_{n{\bf k}}$ to indicate the eigenfunctions of the DFT Kohn-Sham Hamiltonian $\hat{H}_{\rm KS}$, and $\varepsilon_{n\bk}$ the corresponding eigenvalues. This wavefunction is periodic and normalized over the BvK supercell, and can be expressed in the Bloch form:
\begin{equation}
  \psi_{n{\bf k}}({\bf r}) = \frac{1}{\sqrt{\Np}}u_{n{\bf k}}({\bf r})
  e^{i{\bf k}\cdot{\bf r}},
\end{equation}
with $u_{n{\bf k}}$ being a lattice-periodic function normalized in the primitive unit cell of volume $\Omega$. 
The position vector of the atomic nucleus $\kappa$ belonging to the unit cell identified by ${\bf R}_p$ is ${\bm \tau}_{\kappa p} = {\bf R}_p + {\bm \tau}_\kappa$, with ${\bm \tau}_\kappa$ being the position vector within the primitive unit cell. The components of this vector along the Cartesian directions $\alpha = 1,2,3$ are denoted by $\tau_{\kappa\alpha p}$.

Vibrational eigenmodes and eigenfrequencies are obtained by diagonalizing the dynamical matrix, which is computed via DFPT using \texttt{Quantum ESPRESSO}~\cite{giannozzi2017o}. In turn, the dynamical matrix is the Fourier transform of the interatomic force constants $C_{\kappa\alpha p,\kappa'\alpha' p'}$, that is the Hessian of the DFT total energy in the atomic coordinates. The relation between interatomic force constants and dynamical matrix is~\cite{Maradudin1968p}: 
  \begin{equation}\label{eq.dynmat}
  D_{\kappa\alpha,\kappa'\alpha'}({\bf q}) = \frac{1}{\sqrt{M_\kappa M_{\kappa'}}} \sum_p e^{i{\bf q}\cdot{\bf R}_p}
     C_{\kappa\alpha 0,\kappa'\alpha' p} ,
  \end{equation}
where $M_\kappa$ is the mass of the $\kappa$-th nucleus. The eigenvalues and eigenvectors of the dynamical matrix are denoted by $\omega_{{\bf q}\nu}^2$ and $e_{\kappa\alpha,\nu}({\bf q})$,
respectively, and the index $\nu$ runs from 1 to $3M$ where $M$ is the number of atoms in the primitive cell.
$\omega_{{\bf q}\nu}$ corresponds to the vibrational frequency, and $e_{\kappa\alpha,\nu}({\bf q})$ is the
normal mode of vibration or polarization vector.

The electron-phonon matrix element is defined by~\cite{Giustino2007l,Giustino2017z}: \begin{equation}\label{eq:basic4}
      g_{mn\nu}(\bk,\bq) = \< u_{m\bk+\bq} | \Delta_{\bq\nu} v^{\rm KS} | u_{n\bk} \>,
\end{equation}
where the integral is evaluated over the unit cell, and the lattice-periodic component of the variation of the Kohn-Sham potential is given by:
\begin{equation}
    \Delta_{\bq\nu} v^{\rm KS} = \sqrt{\frac{\hbar}{2M_0\w_{\bq\nu}}}\,
   \sum_{\k\a p}
    e^{-i\bq\cdot(\br-\bR_p)} \,\,
   \sqrt{\frac{M_0}{M_\kappa}}
    \,\,
    e_{\k\a,\nu}(\bq)\,
    \frac{\D\, V^{\rm KS}(\br)}
    {\D \tau_{\k\a p}}.
\end{equation}
In this expression, $M_0$ is an arbitrary reference mass introduced for convenience.
MLWFs are defined in terms of Bloch states as~\cite{Marzari2012o}: 
\begin{equation}\label{eq.wannier-def}
  {\rm w}_{m p} ({\bf r}) = {\rm w}_{m 0} ({\bf r}-{\bf R}_p) = \frac{1}{\Np} \sum_{n{\bf k}} e^{i{\bf k}\cdot
    ({\bf r}-{\bf R}_p)}\, U_{nm{\bf k}}\, u_{n{\bf k}}({\bf r}),
\end{equation}
where $U_{nm{\bf k}}$ is a unitary matrix in the indices $m$ and $n$. This matrix is determined by requiring that the functions ${\rm w}_{m p} ({\bf r})$ be as localized as possible~\cite{Pizzi2019s}, i.e.~MLWFs~\cite{Marzari2012o}.

Throughout the manuscript, we use interchangeably the notation
 \begin{equation}
  \frac{1}{\Np}\sum_\bk \qquad \mbox{or} \qquad  \int\frac{d\bk}{\Omega_{\rm BZ}}
 \end{equation}
to indicate summations or integrals over the Brillouin zone. The quantity $\Omega_{\rm BZ}$ on the right indicates the Brillouin zone volume.

\subsection{General principles of Wannier interpolation of electron-phonon matrix elements}
\label{sec.epwtheory}

The localized nature of MLWFs provides the basis for accurate and efficient interpolation of Kohn-Sham wavefunctions and energies~\cite{Yates2007b}. This is achieved by expressing the Kohn-Sham Hamiltonian in the Wannier representation, and noting that the Hamiltonian matrix elements decrease rapidly with the distance between the Wannier function centers. These matrix elements are given by~\cite{Pizzi2019s}:
\begin{equation}\label{eq:basic2}
  H_{mn}({\bf R}_p) = \frac{1}{\Np} \sum_{m'n'{\bf k}} e^{-i{\bf
      k}\cdot{\bf R}_p} U^\dagger_{mm'{\bf k}} H_{m'n'}({\bf k})
  U_{n'n{\bf k}} \,, 
\end{equation}
where $H_{m'n'}({\bf k})$ is the matrix element of the single-particle Kohn-Sham Hamiltonian in the Bloch representation, and the matrix $U_{mn{\bf k}}$ is the same as in Eq.~\eqref{eq.wannier-def}. Once $H_{mn}({\bf R}_p)$ has been determined, Eq.~\eqref{eq:basic2} can be inverted to generate $H_{m'n'}({\bf k})$ anywhere in the Brillouin zone.

Similarly to the electronic case, the dynamical matrix can be expressed in the phonon Wannier representation as~\cite{Giustino2007l}: 
\begin{equation}\label{eq:basic3}
  D_{\kappa\alpha,\kappa'\alpha'}({\bf R}_{p}) = \frac{1}{N_{p}}
  \sum_{{\bf q}\mu\nu} e^{-i{\bf  q}\cdot{\bf R}_{p}} e^\dagger_{\kappa\alpha,\mu}({\bf q}) D_{\mu\nu}({{\bf q}})  e_{\kappa'\alpha',\nu}({\bf q}) \,,
\end{equation}
where $D_{\mu\nu}({{\bf q}})$ is the matrix element of the dynamical
matrix in the Bloch representation for phonons. Once $D_{\kappa\alpha,\kappa'\alpha'}({\bf R}_{p})$ has been determined, this relation can be inverted to obtain the phonon eigenvectors and eigenvalues anywhere in the Brillouin zone; this is a standard procedure employed to compute phonon dispersion relations~\cite{Gianozzi1991l,Gonze1997l}.

Equations~\eqref{eq:basic2} and \eqref{eq:basic3} can be generalized to the case of electron-phonon matrix elements by considering Fourier transforms for both the Kohn-Sham states and for the vibrational eigenmodes. The resulting electron-phonon matrix elements in the Wannier representation are given by~\cite{Giustino2007l}:
\begin{equation}\label{eq:basic5}
  g_{mn\kappa\alpha}({\bf R}_p, {\bf R}_{p'}) = \frac{1}{\Np^2}
  \sum_{{\bf k}, {\bf q}} e^{-i({\bf k}\cdot{\bf R}_p + {\bf
      q}\cdot{\bf R}_{p'})} \sum_{m'n'\nu} \sqrt{\frac{2M_\kappa\omega_{{\bf q}\nu}}{\hbar}}
    e^\dagger_{\kappa\alpha,\nu}({\bf q}) U^\dagger_{mm'{\bf k}+{\bf q}}
    g_{m'n'\nu}({\bf k}, {\bf q}) U_{n'n{\bf k}}\,.
\end{equation}
Once the $g_{mn\kappa\alpha}({\bf R}_p, {\bf R}_{p'})$ have been computed, this relation can be inverted to generate electron-phonon matrix elements anywhere in the Brillouin zone.

\begin{figure}
    \includegraphics[width=\textwidth]{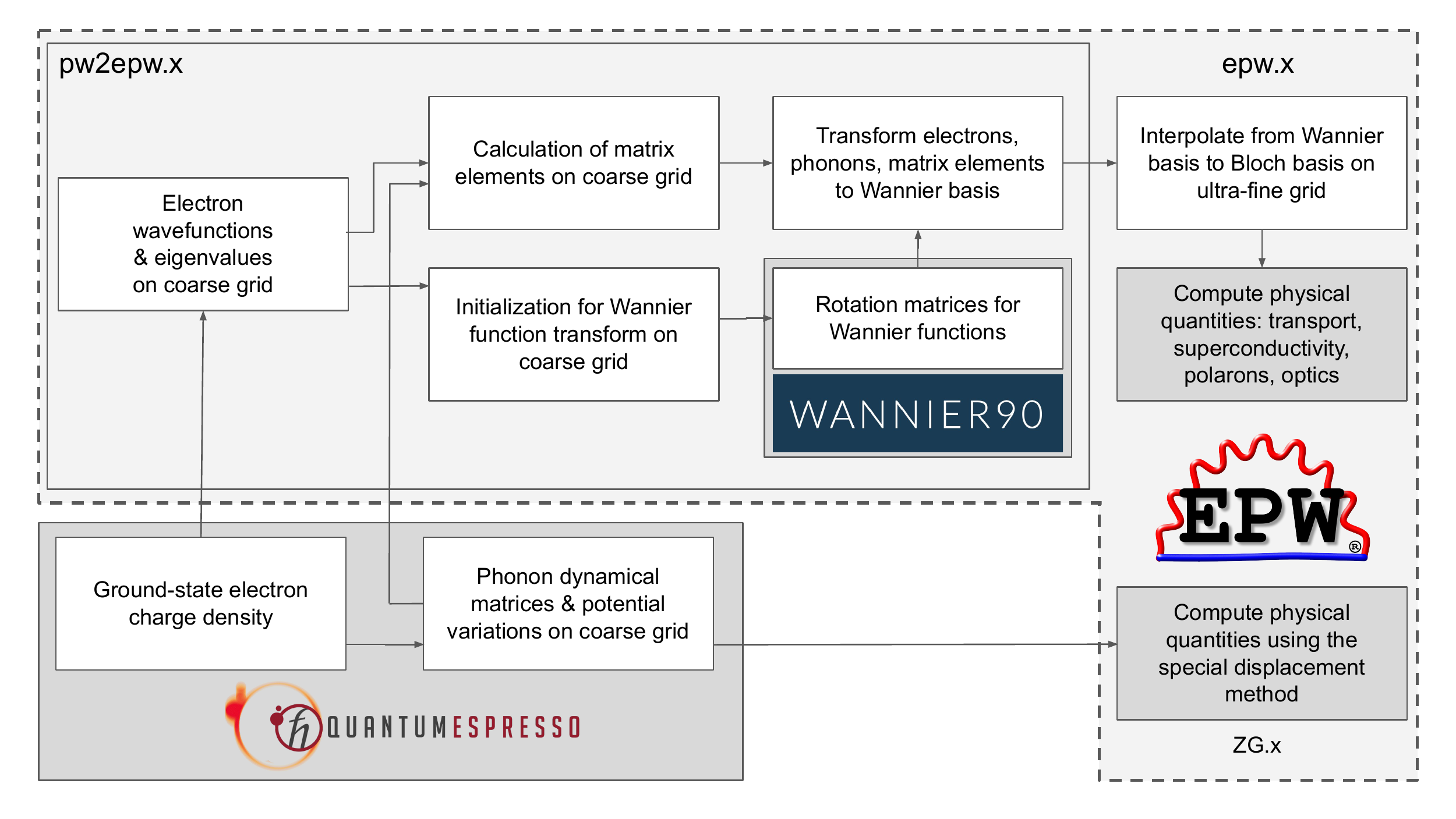}
    \caption{Schematic overview of the interpolation engine of \texttt{EPW} and its relation to the \texttt{Quantum ESPRESSO} and \texttt{Wannier90} codes. The \texttt{EPW/ZG} module is a stand-alone supercell-based code described in Sec.~\ref{sec.sdm}. 
    \label{fig:epw-scheme}}
\end{figure}

Equations~\eqref{eq:basic2}-\eqref{eq:basic5} constitute the backbone of the interpolation engine of \texttt{EPW}. In practice, the interpolation module of \texttt{EPW} reads in the DFT electron density, dynamical matrices, and variations of the Kohn-Sham potential evaluated by \texttt{Quantum ESPRESSO} on a coarse Brillouin zone grid; computes the Kohn-Sham wavefunctions and electron-phonon matrix elements on this grid; calls the \texttt{wannier90} code in library mode to obtain the Wannier matrices in Eq.~\eqref{eq.wannier-def}; transforms Hamiltonian, dynamical matrix, and electron-phonon matrix elements in the Wannier representation; and interpolates all these quantities onto arbitrarily dense $\bk$- and $\bq$-point grids. A qualitative schematic of this process is shown in Fig.~\ref{fig:epw-scheme}.

  \subsection{Treatment of long-range electron-phonon interactions}
\label{sec.longrange}

In semiconductors and insulators, the ionic displacements associated with a phonon can induce a variation of the Kohn-Sham potential that is long-ranged in nature. As a result, the electron-phonon matrix element $g_{mn\nu}(\bk,\bq)$ associated with longitudinal-optical (LO) phonons becomes singular in the long-wavelength limit $\bq\rightarrow 0$. More specifically, in any material exhibiting non-vanishing Born effective charges, the matrix elements associated with LO phonons diverge as $\bq/|\bq|^2$ at small $\bq$. In these cases, the assumption of locality that underpins the methodology described in Sec.~\ref{sec.epwtheory} ceases to hold, and the Wannier interpolation procedure needs to be modified to correctly capture the singularity.

The singularity in the matrix element can be dealt with by considering a multipole expansion of the Kohn-Sham potentials resulting from individual atomic displacements. The first order in this expansion is the dipole potential \cite{Vogl1976l}, which scales as $\bq/|\bq|^2$ and is responsible for the well-known Fr{\"o}hlich electron-phonon interaction \cite{Frohlich1954l}. The modification of the electron-phonon interpolation method to include long-range effects consists of separating short-range and long-range parts in the matrix elements, as follows:
\begin{equation} \label{eq:g_full}
    g_{mn\nu}(\mathbf{k},\mathbf{q}) = 
    g^{\mathcal{S}}_{mn\nu}(\mathbf{k},\mathbf{q}) +
    g^{\mathcal{L},D}_{mn\nu}(\mathbf{k},\mathbf{q})+
    g^{\mathcal{L},Q}_{mn\nu}(\mathbf{k},\mathbf{q}),
\end{equation}
where the first term on the right-hand side is the short-range component, the second term is the dipole component, and the third term is the quadrupole component. Additional multipoles could be considered, but the dipole and quadrupole terms already lead to very accurate results. The dipole matrix element $g^{\mathcal{L},D}_{mn\nu}(\mathbf{k},\mathbf{q})$ was derived in Refs.~\cite{Verdi2015l, Sjakste2015l} and reads:
\begin{eqnarray} \label{eq:frohlich}
    g^{\mathcal{L},D}_{mn\nu}(\mathbf{k},\mathbf{q}) 
    &=&
    i \, \frac{4\pi}{\Omega} \frac{e^{2}}{4\pi\varepsilon_{0}}
    \sum_{\kappa} \left( \frac{\hbar}{2 M_\kappa \omega_{\mathbf{q}\nu}} \right)^{1/2} \nonumber \\
    & & \times \sum_{\mathbf{G}\neq -\mathbf{q}}
    \frac{(\mathbf{q}+\mathbf{G}) \cdot \mathbf{Z_{\kappa}^{*}} \cdot \mathbf{e}_{\kappa\nu}(\mathbf{q})}
    {(\mathbf{q}+\mathbf{G}) \cdot \boldsymbol{\epsilon}^{\infty} \cdot (\mathbf{q}+\mathbf{G})}
    \, \langle \psi_{m\mathbf{k+q}} | e^{i (\mathbf{q}+\mathbf{G}) \cdot \mathbf{r}} 
    | \psi_{n\mathbf{k}} \rangle 
    \, e^{-i (\mathbf{q}+\mathbf{G}) \cdot \boldsymbol{\tau}_{\kappa}} ~.
\end{eqnarray}
In this expression, 
$\varepsilon_{0}$ is the vacuum permittivity, $\boldsymbol{\epsilon}^{\infty}$ is the high-frequency dielectric tensor of the material,
$\mathbf{G}$ represents a reciprocal lattice vector, $\mathbf{Z_{\kappa}^{*}}$ is the Born effective charge tensor of the atom $\kappa$, and the bra-ket indicates the integral over the BvK supercell. This matrix element reduces to the standard Fr{\"o}hlich interaction \cite{Frohlich1954l} when one considers parabolic electron bands and a dispersionless LO mode in a cubic material \cite{Verdi2015l,Sio2022l}. This term is of the order of $|\bq|^{-1}$. The quadrupole contribution
is the second term of the multipole expansion of the Kohn-Sham potential, and is of the order of
$|\bq|^0$. The corresponding matrix element $g^{\mathcal{L},Q}_{mn\nu}(\mathbf{k},\mathbf{q})$ was derived in Refs.~\cite{BruninPRL2020l,BruninPRB2020l,Jhalani2020l,Park2020l} and reads:
\begin{eqnarray} \label{eq:quadrupole}
    & & g^{\mathcal{L},Q}_{mn\nu}(\mathbf{k},\mathbf{q}) 
    =
    \frac{4\pi}{\Omega} \frac{e^{2}}{4\pi\varepsilon_{0}}
    \sum_{\kappa} \left( \frac{\hbar}{2 M_\kappa \omega_{\mathbf{q}\nu}} \right)^{1/2} \nonumber \\
    & & \times \sum_{\mathbf{G}\neq -\mathbf{q}}
    \frac{\frac{1}{2} (\mathbf{q}+\mathbf{G}) \cdot (\mathbf{q}+\mathbf{G}) 
    \cdot  \mathbf{e}_{\kappa\nu}(\mathbf{q}) \cdot \mathbf{Q}_{mn\kappa}}
    {(\mathbf{q}+\mathbf{G}) \cdot \boldsymbol{\epsilon}^{\infty} \cdot (\mathbf{q}+\mathbf{G})}
    \, \langle \psi_{m\mathbf{k+q}} | 
    e^{i (\mathbf{q}+\mathbf{G}) \cdot \mathbf{r}} | \psi_{n\mathbf{k}} \rangle 
    \, e^{-i (\mathbf{q}+\mathbf{G}) \cdot \boldsymbol{\tau}_{\kappa}}.
\end{eqnarray}
In this expression, $\mathbf{Q}_{mn\kappa}$ is the dynamical quadrupole tensor \cite{Royo2019l}. In principle one should add one extra term to this expression, but it was shown that such a term is numerically negligible~\cite{BruninPRL2020l}.

The calculation strategy employed by \texttt{EPW} is as follows. First, the complete matrix elements $g_{mn\nu}(\mathbf{k},\mathbf{q})$ are evaluated on coarse $\mathbf{k}$ and $\mathbf{q}$ grids using DFPT.
Second, the long-range contributions $g^{\mathcal{L},D}_{mn\nu}(\mathbf{k},\mathbf{q})$ and $g^{\mathcal{L},Q}_{mn\nu}(\mathbf{k},\mathbf{q})$ are subtracted from the DFPT matrix elements using Eqs.~\eqref{eq:frohlich} and \eqref{eq:quadrupole}, leaving out the short-range component  $g^{\mathcal{S}}_{mn\nu}(\mathbf{k},\mathbf{q})$ on the coarse grids. Third, the standard Wannier electron-phonon interpolation of Ref.~\cite{Giustino2007l} is applied to the short-range component only.
And fourth, the long-range contributions are added back using Eqs.~\eqref{eq:frohlich} and \eqref{eq:quadrupole} on the fine $\mathbf{k}$ and $\mathbf{q}$ grids.

In this procedure, the overlap integrals between Kohn-Sham wavefunctions appearing in Eqs.~\eqref{eq:frohlich} and \eqref{eq:quadrupole} are evaluated in the $\mathbf{q+G}\rightarrow 0$ limit via \cite{Verdi2015l}:
\begin{equation} \label{eq:overlap}
    \langle \psi_{m\mathbf{k+q}} | 
    e^{i (\mathbf{q}+\mathbf{G}) \cdot \mathbf{r}} | 
    \psi_{n\mathbf{k}} \rangle
    =
    [ U_{\mathbf{k+q}} U_{\mathbf{k}}^{\dagger} ]_{mn} ~,
\end{equation}
where the unitary matrices $U_{mn\mathbf{k}}$ for the Wannier transformation are obtained following the standard procedure \cite{Marzari1997l,Souza2001l}. In the calculation of the quadrupole matrix elements, these overlap integrals should be augmented by an additional Berry-connection term, but numerical tests suggest that this additional contribution is often small~\cite{Ponce2022la,Ponce2022lb}. 

\begin{figure}
    \includegraphics[width=0.7\textwidth]{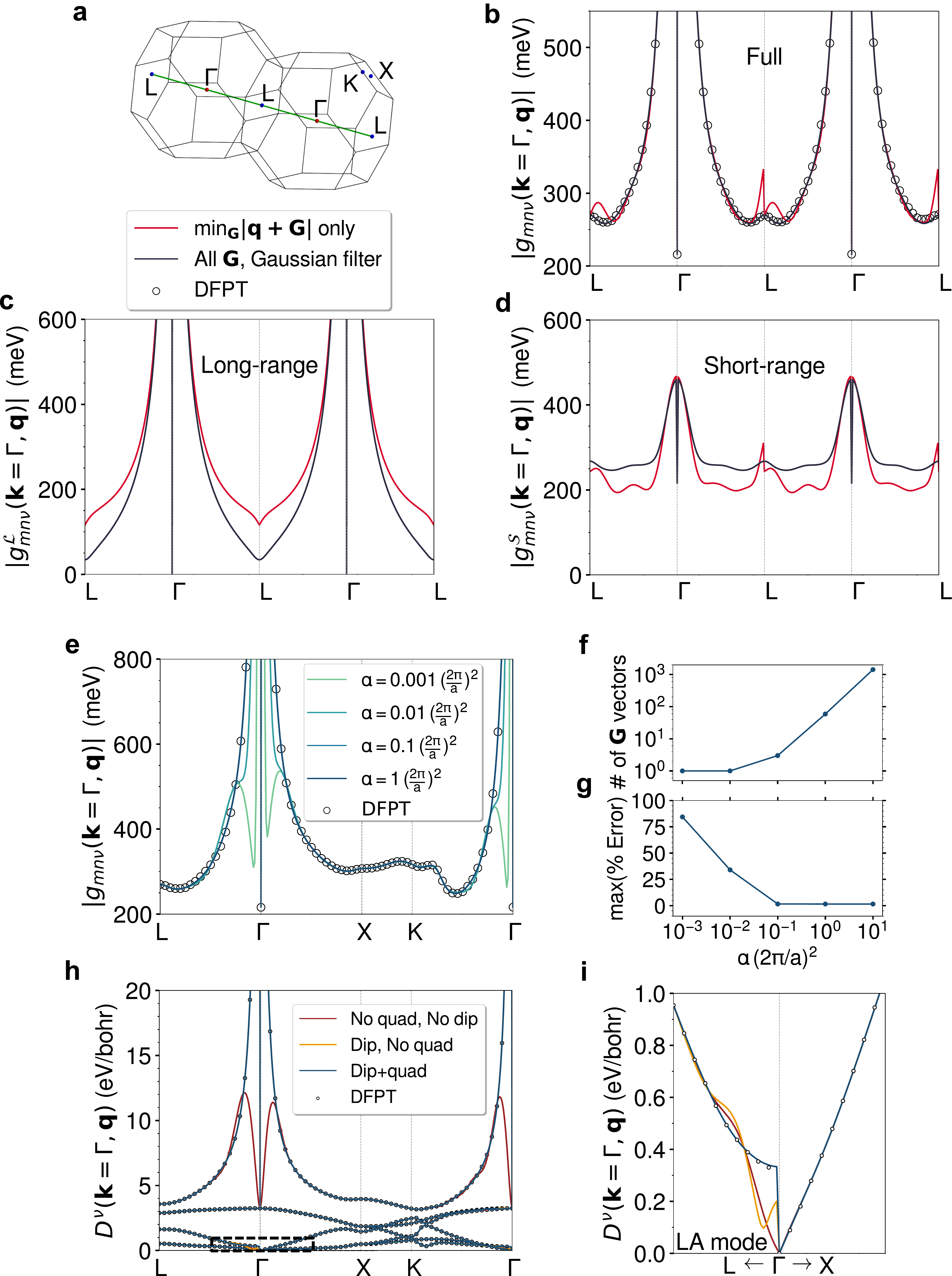}
    \caption{\textbf{Accurate computation of long-range electron-phonon matrix elements with \texttt{EPW}}. (a) Brillouin zone of c-BN, the system used in this example. (b), (c), (d) Full, long-range and short-range electron-phonon matrix elements, respectively, for the LO phonon mode along the $\mathbf{q}$-path shown by the green line in (a). The red lines are for the choice of including only the dominant reciprocal lattice vector in the $\mathbf{G}$ sums. The dark blue lines correspond to the choice of applying a Gaussian filter, the procedure implemented in \texttt{EPW v6}. Circles are the reference data from explicit DFPT calculations. (e) Sensitivity of the interpolation on the $\alpha$ parameter of the Gaussian filter. (f) Number of $\mathbf{G}$ vectors included in the sum as a function of $\alpha$, and (g) corresponding maximum relative error with respect to DFPT reference data. (h) Comparison between the descriptor $D_{\nu}(\mathbf{k}\!=\!0,\mathbf{q})$ obtained from direct DFPT calculations (white circles) and from Wannier interpolation: without including long-range contributions (red line); including the dipole term (orange line); and including both dipole and quadrupole terms (blue line).
    In each case, the corresponding corrections to the phonon dynamical matrix are also included.
    (i) Zoom over the area highlighted in (h), focusing on the LA phonon mode.
    \label{fig_lr}}
\end{figure}

The infinite sum over the $\mathbf{G}$ vectors in Eqs.~\eqref{eq:frohlich} and \eqref{eq:quadrupole}
ensures the periodicity of the matrix elements in reciprocal space. In practical calculations,
this sum might be computationally demanding to converge. In principle, periodicity could be enforced by including only one reciprocal lattice vector in the sum, namely $\tilde{\mathbf{G}}_{\mathbf{q}}$ such that $|\mathbf{q}+\tilde{\mathbf{G}}_{\mathbf{q}}|=\min_{\mathbf{G}}|\mathbf{q+G}|$. However, this choice introduces derivative discontinuities in $g^{\mathcal{L},D}_{mn\nu}(\mathbf{k},\mathbf{q})$ at the Brillouin zone boundaries, which in turn cause spurious oscillations in the interpolation of 
$g^{\mathcal{S}}_{mn\nu}(\mathbf{k},\mathbf{q})$. We illustrate this point in Figs.~\ref{fig_lr}(a)-(d), focusing on the longitudinal-optical (LO) phonon of cubic boron nitride (c-BN).

An alternative strategy to avoid the sum over $\mathbf{G}$ is to cut off the interaction range in Eqs.~\eqref{eq:frohlich} and \eqref{eq:quadrupole} using a Gaussian filter,
$\exp[-(\mathbf{q}+\mathbf{G}) \cdot \boldsymbol{\epsilon}^{\infty}  \cdot (\mathbf{q}+\mathbf{G})/4 \alpha]$. This choice finds motivation in the Ewald summation method that is commonly employed to evaluate the non-analytic contribution to the dynamical matrix  in polar materials~\cite{Gianozzi1991l,Gonze1997l,Baroni2001s}. It was employed in previous versions of \texttt{EPW} as well as in other codes
\cite{Sjakste2015l,BruninPRB2020l,Zhou2021}, and it avoids the derivative discontinuity. However, it does not preserve the periodicity of the matrix elements. In \texttt{EPW v6}, to maintain the periodicity of the matrix elements, we sum over a shell of $\mathbf{G}$ vectors centered around $\tilde{\mathbf{G}}_{\mathbf{q}}$, and then apply the Gaussian filter. Figure~\ref{fig_lr}(c) shows how this procedure yields periodic and smooth long-range matrix elements with the correct behavior next to the singularities, and enables accurate interpolation of the complete matrix elements, as shown in Fig.~\ref{fig_lr}(b).

The $\alpha$ parameter in the Gaussian filter is chosen in such a way as to ensure accurate interpolation (which requires large $\alpha$) whilst using as few $\bG$ vectors as possible in Eqs.~\eqref{eq:frohlich} and \eqref{eq:quadrupole} (which requires small $\alpha$). In Fig.~\ref{fig_lr}(e) we compare the interpolated matrix element $g_{mn\nu}(\mathbf{k},\mathbf{q})$ to explicit DFPT calculations, for different values of $\alpha$, with the reciprocal space summation restricted to those $\mathbf{G}$ vectors such that $(\mathbf{q}+\mathbf{G}) \cdot \boldsymbol{\epsilon}^{\infty} \cdot (\mathbf{q}+\mathbf{G})
< 56\alpha$. With this choice, only $\mathbf{G}$-vectors yielding a value of the Gaussian filter larger than $\exp(-14) \simeq 10^{-6}$ are included in the sum.
We find that, when $\alpha$ is chosen to match the size of the Brillouin zone [$\alpha=1\,(2\pi/a)^{2}$, $a$ is the lattice parameter], good interpolation is achieved. For completeness, in Fig.~\ref{fig_lr}(f) we show how the number of required $\bG$ vectors increases with $\alpha$, 
and in Fig.~\ref{fig_lr}(g) we show the interpolation error in the matrix elements as a function of $\alpha$. We emphasize that, since \texttt{EPW} implements this Gaussian filter, caution should be used when extracting dipolar and quadrupolar matrix elements from \texttt{EPW} for separate post-processing, because $g^{\mathcal{L},D}_{mn\nu}(\mathbf{k},\mathbf{q})$ and $g^{\mathcal{L},Q}_{mn\nu}(\mathbf{k},\mathbf{q})$ have the expected $\bq$-dependence only at long wavelength. 

Figures~\ref{fig_lr}(h) and (i) illustrate the importance of describing dipole and quadrupole interactions using Eqs.~\eqref{eq:frohlich} and \eqref{eq:quadrupole}, for the case of c-BN. To facilitate the comparison between the interpolation results and explicit DFPT calculations for long-wavelength acoustic modes, we use the following descriptor which removes the factor $\omega_{\bq\nu}^{-1/2}$ from the matrix element and averages over electronic degeneracies: 
\begin{equation} \label{eq:deform}
  D_{\nu}(\mathbf{k},\mathbf{q}) = \frac{1}{\hbar N_{\mathrm{W}}}
  \left[2\rho\Omega\hbar\omega_{\mathbf{q}\nu} 
  {\sum}_{mn} |g_{mn\nu}(\mathbf{k},\mathbf{q})|^{2}\right]^{1/2}~.
\end{equation}
Here, the sum over bands is carried over the $N_{\mathrm{W}}$ states in the Wannier manifold and $\rho$ is the mass density of the crystal. $D_{\nu}(\mathbf{k},\mathbf{q})$ has units of energy divided by length, and can be thought of as a deformation potential of sort. In Figs.~\ref{fig_lr}(h) and (i), we choose the manifold composed by the top three valence bands of c-BN, and set $\mathbf{k}=0$. We see that the dipole term is necessary to correctly describe the singular behavior of the LO phonon at long wavelength, and the quadrupole term is necessary to correctly describe the discontinuous behavior of the longitudinal acoustic (LA) phonons at long wavelength [see expanded view in Fig.~\ref{fig_lr}(i)]. The inclusion of both terms guarantees a high-quality interpolation of the electron-phonon matrix elements across the entire Brillouin zone.

The \texttt{EPW} code also implements long-range corrections for the phonon dynamical matrix by computing dipole-dipole, dipole-quadrupole, and quadrupole-quadrupole terms
as discussed in Refs.~\cite{Gianozzi1991l,Gonze1997l,Baroni2001s,Royo2019l}. Systematic tests of the interpolation procedure described in this section are provided in Ref.~\cite{Ponce2021l}.

The expressions for the long-range dipole and quadrupole contributions to the electron-phonon matrix elements given in Eqs.~\eqref{eq:frohlich} and \eqref{eq:quadrupole} are for three-dimensional (3D) bulk crystals. Generalizations of these expressions to the case of two-dimensional (2D) materials and the transition from 3D to 2D have recently been proposed~\cite{Sohier2016l,Deng2021l,Sio2022l,Ponce2022la,Ponce2022lb}.

\section{Capabilities and application examples}\label{sec.functionalities}

In this section we provide an overview of new or expanded capabilities of the \texttt{EPW} code. In particular, we discuss how \texttt{EPW} calculates carrier transport properties via the \textit{ab initio} Boltzmann transport equation (\aibte), including both carrier-phonon (Sec.~\ref{sec.transport}) and carrier-impurity (Sec.~\ref{sec.iiscattering}) scattering; how the superconducting critical temperature and superconducting gap are computed via the solution of the Eliashberg equations (Sec.~\ref{sec.sc}); how we solve the \textit{ab initio} polaron equations to investigate small and large polarons without using large supercells (Sec.~\ref{sec.polarons}); and how we perform calculations of optical absorption spectra in indirect-gap semiconductors by including phonon-assisted optical transitions (Sec.~\ref{sec.indabs}). We also describe an alternative to Wannier interpolation to study electron-phonon interactions, based on the special displacement method (Sec.~\ref{sec.sdm}). Of these features, the superconducting module has been enhanced and expanded with respect to the previous release in 2016~\cite{Ponce2016a}; all the other modules described in the following have been developed \textit{ex novo} since the 2016 release.
  
\subsection{Phonon-limited carrier transport using the \textit{ab initio} Boltzmann transport equation}\label{sec.transport}

\subsubsection{Background and formalism}

The calculation of the electronic transport properties of metals and semiconductors is conveniently dealt with by the \textit{ab initio} Boltzmann transport equation (\textit{ai}BTE). The Boltzmann equation describes the non-equilibrium distribution function of electrons and holes in the presence of external electric or magnetic fields \cite{Ziman1960}. It carries strong predictive power as recently demonstrated for many common semiconductors \cite{Ponce2021l}. Although the Boltzmann formalism is usually derived within the semi-classical approximation, the theory can rigorously be derived from a non-equilibrium many-body Green's function formalism, and is understood as the quasiparticle approximation to the Kadanoff-Baym theory~\cite{Kadanoff1962t,Ponce2020tb,Mahan1987t,Macheda2021t}. 

The \texttt{EPW} code implements the linearized Boltzmann transport equation, which describes the first order response of the distribution function to external fields, and is suitable for investigating the conductivity of metals and the low-field mobility of semiconductors. In this section we focus on the case of phonon-limited transport properties. The extension to include charged defects is discussed in Sec.~\ref{sec.iiscattering}.

We consider homogeneous extended solids held at uniform temperature, and carrier scattering by electron-phonon interactions only, for now. Within the \aibte, the linear response $\partial_{E_{\beta}} f_{n\mathbf{k}}$ of the carrier distribution function to an external electric field $\mathbf{E}$ is given by~\cite{Restrepo2009t,Li2015t,Fiorentini2016t,Zhou2016t,Ponce2018t,Macheda2018t,Sohier2018t,Ma2018t,Ponce2019ta,Ponce2019tb,Ponce2019tc,Lee2020t,Ponce2020ta,BruninPRB2020l,BruninPRL2020l,Ponce2021l}
\begin{multline}\label{eq:iter}
\partial_{E_{\beta}} f_{n\mathbf{k}} = e v_{n\mathbf{k}\beta} \frac{\partial f_{n\mathbf{k}}^0}{\partial \varepsilon_{n\mathbf{k}}} \tau_{n\mathbf{k}} + \frac{2\pi\tau_{n\mathbf{k}}}{\hbar}   \sum_{m\nu} \!\int\! \frac{d\mathbf{q}}{\Omega_{\mathrm{BZ}}} | g_{mn\nu}(\mathbf{k},\mathbf{q})|^2  \partial_{E_{\beta}} f_{m\mathbf{k}+\mathbf{q}} \\
\times \Big[(n_{\mathbf{q}\nu}+1-f_{n\mathbf{k}}^0)\delta(\varepsilon_{n\mathbf{k}} - \varepsilon_{m\mathbf{k+q}}  + \hbar \omega_{\mathbf{q}\nu} ) 
  +  (n_{\mathbf{q} \nu} + f_{n\mathbf{k}}^0)\delta(\varepsilon_{n\mathbf{k}} - \varepsilon_{m\mathbf{k+q}}  - \hbar \omega_{\mathbf{q}\nu}) \Big] .
\end{multline}
In this expression, $f_{n\bk}$ is the electron distribution function, $\partial_{E_{\beta}} f_{n\mathbf{k}}$ is a short-hand notation for $(\partial f_{n\mathbf{k}}/\partial E_\beta)|_{\mathbf{E}=0}$, 
$v_{n\mathbf{k}\alpha} = \hbar^{-1} \partial \varepsilon_{n\mathbf{k}}/\partial k_{\alpha}$ is the intra-band velocity matrix element for the Kohn-Sham eigenvalue $\varepsilon_{n\mathbf{k}}$, and $\delta$ denotes the Dirac delta function. The integral is over the Brillouin zone.
The temperature enters this equation via the Fermi-Dirac and Bose-Einstein equilibrium distribution functions $f^0_{n\mathbf{k}}$ and $n_{\mathbf{q}\nu}$, respectively. 
The quantity $\tau_{n\mathbf{k}}$ in Eq.~\eqref{eq:iter} is the carrier relaxation time, and is obtained from Fermi's golden rule:
\begin{multline}\label{eq:scattering_rate}
  \tau_{n\mathbf{k}}^{-1} = \frac{2\pi}{\hbar} \sum_{m\nu} \!\int\! \frac{d\mathbf{q}}{\Omega_{\mathrm{BZ}}} | g_{mn\nu}(\mathbf{k,q})|^2  \\
  \times  \Big[ (n_{\mathbf{q}\nu} +1 - f_{m\mathbf{k+q}}^0) \delta( \varepsilon_{n\mathbf{k}} - \varepsilon_{m\mathbf{k+q}}   -  \hbar \omega_{\mathbf{q}\nu})  
   +   (n_{\mathbf{q}\nu}  +   f_{m\mathbf{k+q}}^0 )\delta(\varepsilon_{n\mathbf{k}} - \varepsilon_{m\mathbf{k+q}}  +  \hbar \omega_{\mathbf{q}\nu}) \Big].
\end{multline}
The electrical conductivity tensor $\sigma_{\alpha\beta}$, which corresponds to the variation of the current density with respect to the electric field, is computed from $\partial_{E_{\beta}} f_{n\mathbf{k}}$ using:
\begin{equation}\label{eq:sigma}
\sigma_{\alpha\beta} = -\frac{e}{\Omega} \sum_n \int \frac{d\mathbf{k}}{\Omega_{\mathrm{BZ}}} \, v_{n\mathbf{k}\alpha}  \partial_{E_{\beta}}f_{n\mathbf{k}},
\end{equation}
and the drift mobility is obtained from this expression upon dividing by the carrier density~$n_{\rm c}$:
\begin{equation}
\mu_{\alpha\beta} = \frac{\sigma_{\alpha\beta}}{e\,n_{\mathrm{c}}}.
\label{eq:mu}
\end{equation}
These equations are valid for metals and for electrons or holes in semiconductors taken separately. A common approximation to the \textit{ai}BTE is the relaxation time approximation, which consists of neglecting the second term on the right-hand side of Eq.~\eqref{eq:iter}. In this case the mobility takes the simpler form: \begin{equation}\label{eq:serta}
\mu_{\alpha\beta}^{\rm SERTA} = -\frac{e}{n_{\mathrm{c}}\Omega}\sum_n \int \frac{d\mathbf{k}}{\Omega_{\mathrm{BZ}}} \frac{\partial f_{n\mathbf{k}}^0}{\partial \varepsilon_{n\mathbf{k}}} v_{n\mathbf{k}\alpha} v_{n\mathbf{k}\beta} \tau_{n\mathbf{k}}.
\end{equation}
We refer to this approximation as the ``self-energy relaxation time approximation'' (SERTA) since the transport lifetime in Eq.~\eqref{eq:scattering_rate} corresponds to the quasiparticle lifetime, which is proportional to the imaginary part of the electron-phonon self-energy \cite{Ponce2021l,Giustino2017z,Ponce2020tb}. 

The electrical conductivity and the drift mobility described by Eq.~\eqref{eq:iter} can be measured by time-of-flight measurements or by THz photo-conductivity measurements. In the case of Hall and De Haas–Van Alphen measurements, an additional magnetic field $\mathbf{B}$ is applied, and the resulting Lorentz force must be taken into account in the \aibte. In these cases, \texttt{EPW} solves the following \aibte\ equation~\cite{Macheda2018t,Ponce2020tb,Macheda2020t,Ponce2021l}: 
\begin{multline}\label{eq:iterwithbimpl}
 \Big[ 1 - \frac{e}{\hbar}\tau_{n\mathbf{k}} ({\textbf{v}}_{n\mathbf{k}} \times {\textbf{B}}) \cdot \nabla_{\mathbf{k}}
\Big]\partial_{E_{\beta}} f_{n\mathbf{k}}(\mathbf{B}) = e v_{n\mathbf{k}\beta} \frac{\partial f_{n\mathbf{k}}^0}{\partial \varepsilon_{n\mathbf{k}}} \tau_{n\mathbf{k}} \\
+ \frac{2\pi\tau_{n\mathbf{k}}}{\hbar}
  \sum_{m\nu} \!\int\! \frac{d\mathbf{q}}{\Omega_{\mathrm{BZ}}} | g_{mn\nu}(\mathbf{k},\mathbf{q})|^2  \partial_{E_{\beta}} f_{m\mathbf{k}+\mathbf{q}}(\mathbf{B}) \\
 \times  \Big[(n_{\mathbf{q}\nu}+1-f_{n\mathbf{k}}^0)\delta(\varepsilon_{n\mathbf{k}} - \varepsilon_{m\mathbf{k+q}}  + \hbar \omega_{\mathbf{q}\nu} )  
  +  (n_{\mathbf{q} \nu} + f_{n\mathbf{k}}^0)\delta(\varepsilon_{n\mathbf{k}} - \varepsilon_{m\mathbf{k+q}}  - \hbar \omega_{\mathbf{q}\nu}) \Big].
\end{multline}
The variation $\partial_{E_{\beta}} f_{n\mathbf{k}}(\mathbf{B})$ is computed for magnetic fields sufficiently small that their effect on the electronic and vibrational properties can be ignored; accordingly, the Kohn-Sham energies, phonons, and their couplings in the above equation are all evaluated for $\mathbf{B}=0$. 
The application of a magnetic field in the direction $\hat{\mathbf{B}}$ results 
in an orthogonal flow of the charge carriers which can be described by the linear response of the mobility to the field:
\begin{equation}\label{eq:muH}
\mu_{\alpha\beta}^{(2)}(\hat{\mathbf{B}}) = -\lim_{\mathbf{B}\rightarrow \mathbf{0}}\frac{1}{|\mathbf{B}|} \bigg[  \frac{1}{n_{\mathrm{c}} \Omega} \sum_n \int \frac{d\mathbf{k}   }{\Omega_{\mathrm{BZ}}} v_{n\mathbf{k}\alpha}  \partial_{E_{\beta}}f_{n\mathbf{k}}(\mathbf{B}) - \mu_{\alpha\beta} \bigg].
\end{equation}
From this expression, we obtain the Hall mobility in terms of the drift mobility as follows:
\begin{equation}\label{eq:hallmob}
  \mu^{\mathrm{H}}_{\alpha\beta}(\hat{\mathbf{B}}) = \sum_\gamma r^{\rm H}_{\alpha\gamma}(\hat{\mathbf{B}})\mu_{\gamma\beta}~,
\end{equation}
where we have introduced the dimensionless Hall factor as~\cite{Ponce2021l}:
\begin{equation}\label{eq:hallfac}
 r^{\rm H}_{\alpha\gamma}(\hat{\mathbf{B}}) = \sum_{\gamma\delta} \mu_{\alpha\gamma}^{-1}  \mu^{(2)}_{\gamma\delta}(\hat{\mathbf{B}})  \mu_{\delta\beta}^{-1}. 
\end{equation}
Also in this case, one can simplify the solution of Eq.~\eqref{eq:iterwithbimpl} by ignoring the second term on the right-hand side. This choice leads to the SERTA approximation in the presence of a magnetic field.

\subsubsection{Computational considerations}

\texttt{EPW} implements an iterative solver to obtain $\partial_{E_{\beta}} f_{n\mathbf{k}}$ from Eq.~\eqref{eq:iter}, with the possibility of Broyden mixing~\cite{Broyden1965t} to accelerate the convergence.  The accurate evaluation of Eqs.~\eqref{eq:sigma} and \eqref{eq:mu} is computationally  demanding because it requires the knowledge of $\partial_{E_\beta} f_{n\mathbf{k}}$ for a dense set of $\bf k$-points in an energy window of the order of a few tens of meV around the Fermi energy. In particular, since the scattering integral on the right-hand side of Eq.~\eqref{eq:iter} couples linear response coefficients at $\mathbf{k}$ and $\mathbf{k+q}$ points, the $\bk$-grid and the $\bq$-grid must be commensurate. As this term is not evaluated in the SERTA approximation, SERTA calculations can instead be performed using incommensurate grids, including for example random and quasi-random sampling~\cite{Ponce2018t}. It should be noted, however, that the SERTA approximation tends to underestimate transport coefficients by up to 50\% as compared to the \aibte~\cite{Ponce2021l}. 

Regardless of the approximation chosen to calculate transport coefficients, \texttt{EPW} exploits crystal symmetry operations to eliminate symmetry-equivalent wavevectors in the solution of Eq.~\eqref{eq:iter}; accordingly, $\partial_{E_\beta} f_{n\mathbf{k}}$ is evaluated within the irreducible wedge of the Brillouin zone. Furthermore, the solution of Eq.~\eqref{eq:iter} is restricted to the set of wavevectors for which both the initial and the final electronic state lie within a user-defined energy window around a reference energy. In particular, the wavevectors $\bk$ and $\bq$ are retained only if there exists a pair of bands $m$ and $n$ such that $|\ve_{n\mathbf{k}}-E^{\rm ref}|<\Delta$ and $|\ve_{m\mathbf{k+q}}-E^{\rm ref}|<\Delta$, where $E^{\rm ref}$ is a reference energy and $\Delta$ is the width of the window. In the case of metals, the reference energy is set to the Fermi energy. In the case of semiconductors, the Fermi energy is calculated from the user-specified temperature and carrier concentration, using the bisection method.

Magneto-transport calculations using Eq.~\eqref{eq:iterwithbimpl} are harder to converge than calculations without magnetic fields. The iterative solution is initialized by using the result of Eq.~\eqref{eq:iter} as a first approximation for $\partial_{E_{\beta}} f_{n\mathbf{k}}(\mathbf{B})$, and the quantity $\nabla_{\mathbf{k}} \partial_{E_{\beta}} f_{n\mathbf{k}}(\mathbf{B})$ appearing on the left-hand side of Eq.~\eqref{eq:iterwithbimpl} is evaluated via finite differences. The correctness of the solution is tested by checking for the conservation of the carrier density, which corresponds to the condition 
$\sum_{n\mathbf{k}} \partial_{E_{\beta}} f_{n\mathbf{k}}(\mathbf{B}) = 0$.

To ensure numerically-accurate evaluation of transport coefficients with \texttt{EPW}, 
it is important to make sure that results be converged with respect to (i) the density of the Brillouin zone grids employed in Eq.~\eqref{eq:iter} and Eq.~\eqref{eq:iterwithbimpl}; (ii) the density of the coarse grids used for the Wannier-Fourier interpolation of the DFT and DFPT data~\cite{Ponce2021l}; (iii) the Gaussian smearing employed to compute the Dirac delta functions appearing in Eq.~\eqref{eq:iter}; and (iv) the sensitivity of the results to lattice parameters, exchange and correlation functionals, and pseudopotentials. To facilitate the convergence with respect to the Gaussian smearing and the $\bk$-point sampling, \texttt{EPW} offers the possibility of using the adaptive smearing method of Ref.~\cite{Li2014t}.

\subsubsection{Application example}

To demonstrate the implementation of the transport module in \texttt{EPW}, we investigate the electron and hole mobilities of cubic boron nitride, as well as the associated Hall factors.
We use the relativistic Perdew-Burke-Ernzerhof (PBE) parametrization~\cite{Perdew1996p} of the generalized gradient approximation to DFT. The pseudopotentials are norm-conserving, generated using the ONCVPSP code~\cite{Hamann2013s}, and optimized via the PseudoDojo initiative~\cite{setten2018p}.
We consider room temperature and low carrier concentrations of 10$^{\rm 13}$~cm$^{-3}$. c-BN is a polar wide gap semiconductor with isotropic Born effective charges $Z^*_{\rm B} = 1.91~e$ and $Z^*_{\rm N} = -Z^*_{\rm B}$; isotropic dynamical quadrupoles $Q_{\rm B}=3.46 \,ea_0$ and $Q_{\rm N}=-0.63\, ea_0$, where $a_0$ is the Bohr radius; and an isotropic high-frequency relative dielectric constant $\epsilon^{\infty}=4.54$~\cite{Ponce2021l}. As already discussed in Fig.~\ref{fig_lr}, long-range electron-phonon interactions are important in the case of c-BN. Wannier functions are calculated separately for the valence and the conduction band manifold in order to reduce the computational cost.

\begin{figure*}
\includegraphics[width=0.9\columnwidth]{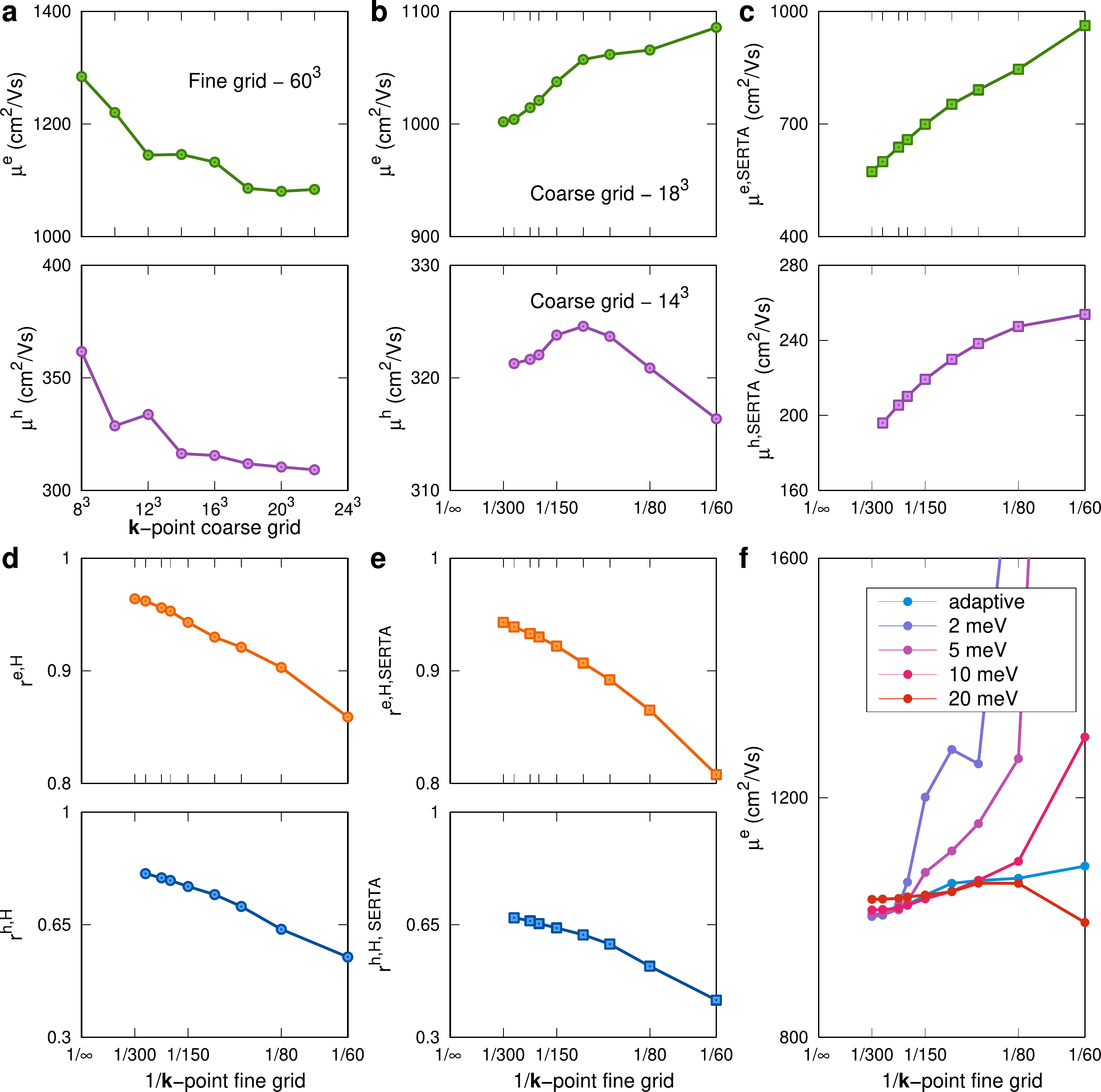}
\caption{\textbf{Calculations of phonon-limited carrier mobility using \texttt{EPW}.} (a) Convergence test for the electron and hole mobilities of c-BN with respect to the number of $\bk$-points on the coarse grid. The number of $\bq$-points is set to half the number of $\bk$-points. (b) Convergence of mobilities w.r.t. the number of grid points on the fine grids. The number of $\bq$-points is set to be the same as the number of $\bk$-points. (c) Same as in (b), but this time for calculations within the SERTA approximation.
(d) and (e): Hall factors of c-BN for electrons and holes, as a function of the number of points on the fine grids,
for the full \aibte\ and the SERTA approximation, respectively. The number of $\bq$-points is set to be the same as the number of $\bk$-points. (f) Dependence of the electron drift mobility of c-BN with the Gaussian smearing.}
\label{fig:transport}
\end{figure*}

Figure~\ref{fig:transport}(a) shows the convergence of calculated mobilities with the coarse grid of $\mathbf{k}$-points used in the Wannierization procedure. For this test, we employ a coarse $\mathbf{q}$-point grid with half the number of the $\mathbf{k}$-points; and we employ identical fine grids with $50^3$ $\mathbf{k}$- and $\mathbf{q}$-points for the solution of Eq.~\eqref{eq:iter}; we use an energy window $\Delta=0.3$~eV around the reference energy $E^{\rm ref}$ set to the band edge, and we employ adaptive smearing. This panel shows that the mobilities are converged with an accuracy of $1\%$ when coarse grids with $18^3$ and $14^3$ points are employed for electrons and holes, respectively.

Figure~\ref{fig:transport}(b) shows the convergence of calculated mobilities with respect to the fine grids. We see that both electron and hole mobilities are almost converged for grids consisting of $250^3$ points. 

In Fig.~\ref{fig:transport}(c) we show convergence curves for the mobility evaluated in the SERTA approximation. In this case, the convergence with the density of points in the fine grids is much slower than for complete \aibte\ calculations. More importantly, SERTA mobilities can differ by up to a factor of two from the corresponding \aibte\ results. The large difference between full \aibte\ calculations and SERTA seems to be a common trend in polar materials~\cite{Ponce2021l}. For the same $\mathbf{k}$- and $\mathbf{q}$-point grids, the computational saving afforded by the SERTA approximation is minimal; therefore, we recommend using SERTA only when the use of commensurate grids is too demanding, for example in the case of systems with many atoms in the unit cell. 

Figures \ref{fig:transport}(d) and (e) show the computed Hall factors for electrons and holes in c-BN, respectively. The Hall factor in the SERTA approximation appears to be close to the full \aibte\ result, which is expected since it is defined as the ratio of two mobilities, see Eq.\eqref{eq:hallfac}~\cite{Paola2020t}. 

In Fig.~\ref{fig:transport}(f) we analyze the role of the Gaussian smearing parameter in the calculations. To this end, we compute the electron mobility of c-BN as a function of grid size, for varying smearing parameter. It is seen that the size of the grid necessary for convergence increases when decreasing the smearing in the few meV's range. On the other hand, the use of adaptive smearing affords fast convergence even for the smaller grid sizes. Based on this comparison, adaptive smearing is the computationally most convenient strategy.

\subsection{Defect-limited carrier transport using the \textit{ab initio} Boltzmann transport equation}\label{sec.iiscattering}

\subsubsection{Background and formalism}

Electron-phonon interactions are the dominant scattering mechanism in high-purity single crystals with low defect concentration, typically up to $10^{15}$ cm$^{-3}$~\cite{Ashcroft1976i}. In tetrahedral semiconductors and a host of other materials, charged defects are ubiquitous since dopant elements are employed to introduce free carriers~\cite{Lundstrom2000i}. Donors release electrons to the conduction band, thus becoming positively-charged defects; similarly, acceptors release holes into the valence band, and become negatively-charged defects. In either case, the impurity generates a long-ranged Coulomb potential that scatters charge carriers. This scattering mechanism tends to dominate over electron-phonon processes at high doping concentrations.

There exist popular semi-empirical relations to estimate the effect of ionized impurity scattering on carrier transport in semiconductors, such as the Brooks-Herring~\cite{Brooks1955i} and Conwell-Weisskopf~\cite{Conwell1954i} formulas. However, these expressions rely on simplified parabolic band models and do not carry predictive power in the case of materials with multi-band or multi-valley band extrema. First-principles calculations offer a modern alternative to these earlier approaches, and have achieved considerable success in recent years~\cite{Restrepo2009i,Graziosi2020i,Ganose2021i,Lu2022i}. The \texttt{EPW} code implements a module for charged defect scattering based on a randomized distribution of point charges. This approach is described in detail in Ref.~\cite{Leveillee:2023} and summarized below.

For convenience we rewrite the \aibte\ from Eq.~\eqref{eq:iter} in the more compact form:
\begin{equation}\label{eq:BTE1}
  -e  v_{n\mathbf{k}\beta}\frac{\partial f^0_{n\mathbf{k}}}{\partial \epsilon_{n\mathbf{k}}}
  = \sum_{m}\int\frac{d\bq}{\Omega_{\rm BZ}} \left( \tau^{-1}_{m\mathbf{k}+\mathbf{q} \to n\mathbf{k}} \, \partial_{E_{\beta}}f_{m\mathbf{k}+\mathbf{q}} - \tau^{-1}_{n\mathbf{k} \to m\mathbf{k}+\mathbf{q}} \, \partial_{E_{\beta}}f_{n\mathbf{k}} \right),
\end{equation}
where the quantity $\tau^{-1}_{n\mathbf{k} \to m\mathbf{k}+\mathbf{q}}$ is the partial scattering rate from the Kohn-Sham state $n\bk$ to the state $m\bk+\bq$. In the case of electron-phonon scattering, $\tau^{-1}_{n\mathbf{k} \to m\mathbf{k}+\mathbf{q}}$ is obtained from Eq.~\eqref{eq:scattering_rate} by removing the summation over $m$ and the integral over the Brillouin zone, i.e. $\sum_m\int d\bq/\Omega_{\rm BZ}$. 
When both electron-phonon scattering and charged impurity scattering are taken into account, the partial scattering rate is written as the sum of the individual partial rates
\begin{equation}\label{eq:totrate}
  \displaystyle
  \frac{1}{\tau_{n\mathbf{k} \to m\mathbf{k}+\mathbf{q}}} = 
  \frac{1}{\tau^{\rm ph}_{n\mathbf{k} \to m\mathbf{k}+\mathbf{q}}} + 
  \frac{1}{\tau^{\rm imp}_{n\mathbf{k} \to m\mathbf{k}+\mathbf{q}}},
\end{equation}
where the superscripts refer to carrier-phonon (ph) and carrier-impurity (imp) scattering, respectively.
In \texttt{EPW}, the carrier-impurity partial scattering rate $1/\tau^{\rm imp}_{n\bk \to m\bk+\bq}$ is calculated under the following simplifying approximations: (i) each impurity is described by an idealized point charge, embedded in the dielectric continuum of the host material; (ii) the scattering rate from each impurity is evaluated within the first Born approximation; (iii) defects are sufficiently diluted that the scattering rates from different impurities are additive; (iv) impurities are randomly distributed, and this random distribution is formally taken into account by using the Kohn and Luttinger ensemble average~\cite{Kohn1957i}. Within these approximations, the charged impurity scattering rate reads~\cite{Leveillee:2023}:
\begin{equation}\label{eq:tauave3}
  \frac{1}{\tau^{\text{imp}}_{n\bk\rightarrow m\bk+\bq}} = 
  N_{\rm imp}
  \frac{2\pi}{\hbar} \left[\frac{e^2}{4\pi\varepsilon_0} \frac{4\pi Z}{\Omega}\right]^2
  \sum_{\bf G \ne -{\bf q}}
  \frac{\,|\langle \psi_{m\mathbf{k+q}} | 
    e^{i (\mathbf{q}+\mathbf{G}) \cdot \mathbf{r}} | 
    \psi_{n\mathbf{k}} \rangle|^2}{|(\bq+\bG)\cdot\!\boldsymbol{\epsilon}^0\!\cdot(\bq+\bG)|^2}
  \delta(\varepsilon_{n\bk}-\varepsilon_{m\bk+\bq}),
\end{equation}
where $N_{\rm imp}$ is the number of impurities per crystal unit cell (dimensionless), $Ze$ is the charge of each impurity, and $\boldsymbol{\epsilon}^0$ is the static relative dielectric constant tensor.
The scattering rate in Eq.~\eqref{eq:tauave3} takes into account the DFT electronic band structure and the lattice screening, including possible dielectric anisotropy. The main advantage of this model as compared to explicit calculations of charged defects in supercells is that it can be used systematically without requiring detailed knowledge of the defect physics and energetics in each material.

\subsubsection{Computational considerations}

In the \texttt{EPW} code, the scattering rate given in Eq.~\eqref{eq:tauave3} is added to the electron-phonon scattering rate using Eqs.~\eqref{eq:totrate} and \eqref{eq:BTE1}, and the \aibte\ is solved as already described in Sec.~\ref{sec.transport}. The computational overhead as compared to phonon-only calculations is minimal.

In Eq.~\eqref{eq:tauave3}, the sum over $\bG$-vectors is handled via the Gaussian filter described in Sec.~\ref{sec.longrange}, and the same considerations apply here. The overlap integrals between the initial and final Kohn-Sham states appearing in Eq.~\eqref{eq:tauave3}, $\langle \psi_{m\mathbf{k+q}} |  e^{i (\mathbf{q}+\mathbf{G}) \cdot \mathbf{r}} | \psi_{n\mathbf{k}} \rangle$, are evaluated by means of the unitary Wannier function matrices following Eq.~\eqref{eq:overlap}.

The integral over the scattering wavevectors $\bq$ in Eq.~\eqref{eq:BTE1} of the impurity scattering 
rates given in Eq.~\eqref{eq:tauave3} contains a singular $|\bq|^{-4}$ term that is not integrable. This issue is resolved by introducing the screening of the defect potential by free carriers released upon ionization. To this end, we replace $\boldsymbol{\epsilon}^0$ in Eq.~\eqref{eq:tauave3} by the total dielectric function:
  \begin{equation}
      \boldsymbol{\epsilon}^0_{\rm tot} = \boldsymbol{\epsilon}^0 + \frac{q_{\rm TF}^2}{q^2}\boldsymbol{I},
  \end{equation}
where $\boldsymbol{I}$ denotes the identity matrix, and $q_{\rm TF}$ is the Thomas-Fermi wavenumber obtained from the long-wavelength limit of the Lindhard function~\cite{Ashcroft1976i,Lindhard1954i,Lu2022i}:  
  \begin{equation}\label{eq:qtf}
    q_{\rm TF}^2 = \frac{e^2}{4\pi\varepsilon_0}\frac{ 4\pi}{\Omega} 
       2\sum_{n}\int \frac{d\bk}{\Omega_{\rm BZ}} \left|\frac{\partial f^0_{n\bk}}{\partial\varepsilon_{n\bk}}\right|.
  \end{equation}
Temperature enters this expression via the equilibrium Fermi-Dirac distribution of the electrons or holes, $f_{n\bk}^0$.

The concentration of charged defects is an external input parameter in these calculations. In the case of ionized impurities in semiconductors, it is also possible to use a simple estimate for the fraction $f_{\rm ii}$ of ionized impurities at a given temperature by using the impurity energy level in the gap $\varepsilon_{\rm d}$~\cite{Ashcroft1976i,Sanders2021i,Lu2022i}. 
\begin{equation}
f^2_{\rm ii} = \max\left(1, \frac{1}{n_d \,\Omega}\sum_{n}\int\frac{d\bk}{\Omega_{\rm BZ}} 
    f(\varepsilon_{n\mathbf k}-\varepsilon_{\rm d})\right)~,
\end{equation}
where $f$ is the Fermi-Dirac distribution and $n_d$ is the total concentration of defects that may thermally ionize. 

\subsubsection{Application example}

To illustrate the impurity scattering capability of \texttt{EPW}, we calculate the mobility of electrons and holes in silicon as a function of both temperature and dopant concentration. We consider coarse Brillouin zone grids with 12$^3$ $\bk$-points and 6$^{3}$ $\bq$-points, and fine grids with 100$^3$ points. We use the PBE functional \cite{Perdew1996p} and ONCV pseudopotentials~\cite{Hamann2013s,Schlipf2015s}, and we include spin-orbit coupling for the valence bands. We account for quadrupole corrections using the procedure described in Sec.~\ref{sec.longrange}.

\begin{figure*}
\includegraphics[width=0.9\textwidth]{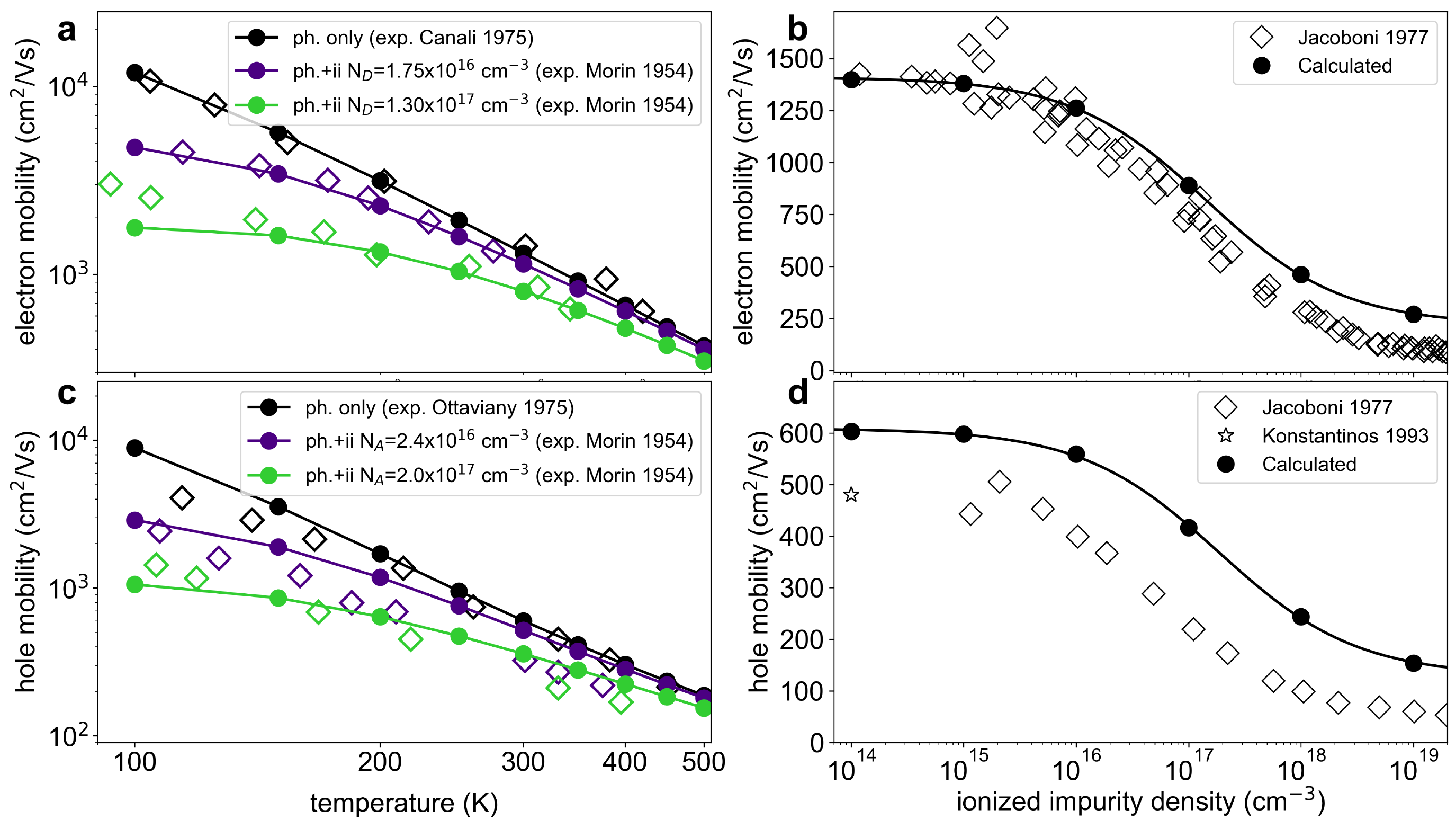}
\caption{\textbf{Calculations of phonon- and impurity-limited carrier mobility using \texttt{EPW}.} (a) Mobility of silicon as a function of temperature, for the following three cases: phonon scattering only (black disks and lines); phonon and ionized impurity scattering, with an impurity concentration of $1.75\cdot10^{16}$ cm$^{-3}$ (indigo disks and lines), and with a concentration of $1.3\cdot 10^{17}$ cm$^{-3}$ (green disks and lines). Experimental data for the same concentrations are shown as diamonds of the same color~\cite{Morin1954i,Canali1975i,Ottaviani1975i}. (b) Same as in (a) but for the hole mobility of silicon. The impurity concentrations are $2.4\cdot10^{16}$ cm$^{-3}$ (indigo) and $2.0\cdot10^{17}$ cm$^{-3}$ (green), respectively. (c) Electron mobility of silicon at 300~K, as a function of ionized impurity concentration: black disks and lines are calculations, black diamonds are experiments~\cite{Jacobini1977i,Konstantinos1993i}. (d) Same as in (c) but for the hole mobility.} 
\label{fig:ii_si}
\end{figure*}

Figure~\ref{fig:ii_si}(a) shows the calculated electron mobility of silicon as a function of temperature. When considering phonon scattering only, our calculations are in very good with measurements on high-purity silicon~\cite{Canali1975i}. Upon introducing ionized impurity scattering, the electron mobility at 100~K decreases from 11,813 cm$^{2}$/Vs to 4,725 cm$^{2}$/Vs for a dopant concentration of 1.75$\cdot$10$^{16}$ cm$^{-3}$, and to 1,769 cm$^{2}$/Vs for dopant concentration of 1.3$\cdot$10$^{17}$ cm$^{-3}$. These calculations agree well with experimental data \cite{Morin1954i}. At higher temperatures, the reduction in the mobility due to impurity scattering is less significant since phonons provide the dominant scattering mechanism. The hole mobility in Fig.~\ref{fig:ii_si}(b) exhibits a similar trend. Upon introducing ionized impurity scattering, the hole mobility at 100~K decreases from 8,877 cm$^{2}$/Vs to 2,884 cm$^{2}$/Vs for a dopant concentration of 2.4$\cdot$10$^{16}$ cm$^{-3}$, and to 1,056 cm$^{2}$/Vs for a dopant concentration of 2.0$\cdot$10$^{17}$ cm$^{-3}$. Also in this case, the agreement with experimental data is very good \cite{Morin1954i}.

Figure~\ref{fig:ii_si}(c) shows the calculated electron mobility of silicon at 300~K, as a function of ionized impurity concentration. Up to a dopant concentration of 10$^{16}$~cm$^{-3}$, phonon scattering dominates and the mobility is relatively insensitive to impurity scattering. Beyond this concentration, the mobility decreases sharply and approximately as the inverse of the impurity density, in line with Eq.~\eqref{eq:tauave3}. The calculations agree well with experiments up to an impurity concentration around 10$^{18}$ cm$^{-3}$ \cite{Jacobini1977i}. Beyond this concentration, it is expected that additional mechanisms such as two-impurity scattering and plasmon scattering~\cite{Caruso2016i} will further reduce the mobility. The hole mobility, which is shown in Fig.~\ref{fig:ii_si}(d), follows a similar trend. The calculated hole mobility at low doping, 603 cm$^{2}$/Vs, slightly overestimates the experimental range 450-500 cm$^{2}$/Vs; this effect can be traced to the underestimation of the heavy hole mass by DFT ~\cite{Ponce2018t}. Upon increasing the impurity concentration, the mobility decreases following the same trend as for the electrons. These calculations agree well with experiments \cite{Jacobini1977i,Konstantinos1993i} and with previous first-principles calculations~\cite{Lu2022i}.

  \subsection{Phonon-mediated superconductivity using the \textit{ab initio} Eliashberg theory}\label{sec.sc}

\subsubsection{Background and formalism}

First-principles calculations of phonon-mediated superconductors are primarily based on three approaches, namely  semi-empirical methods based on the McMillan formula~\cite{McMillan1968s}, the \textit{ab initio} Eliashberg theory~\cite{Eliashberg1960s,Eliashberg1961s}, and the density-functional theory for superconductors~\cite{Oliveira1988s,Luders2005s,Marques2005s,Sanna2020s}. The \texttt{EPW} code implements the former two approaches. In this section, we briefly review methods based on the McMillan formula, and then we describe the Eliashberg formalism and its implementation.

The superconducting critical temperature $T_\mathrm{c}$ can be estimated using standard semi-empirical formulas. In this case, the \texttt{EPW} code allows the user to calculate the Allen-Dynes formula for strong-coupling superconductors~\cite{Allen1975s}:
  \begin{equation}\label{xsceq1}
    \kt_\mathrm{c}^\mathrm{AD}=\frac{\hbar\omega_{\log }}{1.2} 
    \exp \left[\frac{-1.04(1+\lambda)}{\lambda-\mu_\mathrm{c}^*(1+0.62 \lambda)}\right],
  \end{equation}
where $k_{\rm B}$ is Boltzmann's constant, $\mu_\mathrm{c}^*$ is the semi-empirical Coulomb pseudopotential, $\omega_\mathrm{log}$ is the logarithmic average of the phonon frequencies, and $\lambda$ is the electron-phonon coupling constant, as defined in Ref.~\cite{Giustino2017z}. In addition, the code allows the user to calculate the critical temperature using a more recent prescription based on a machine learning approach~\cite{Xie2022s}:
  \begin{equation}\label{xsceq2}
    T_\mathrm{c}^\mathrm{ML}=f_{\omega}f_{\mu}T_\mathrm{c}^\mathrm{AD},
\end{equation}
where the correction factors $f_\omega$ and $f_\mu$ are given by:
\begin{equation}
  f_{\omega}=1.92\frac{\lambda+\omega_\mathrm{log}/ \bar{\omega}_{2}
  -(\mu_\mathrm{c}^{*})^{1/3}}
  {\lambda^{1/2}\exp(\omega_\mathrm{log} / \bar{\omega}_{2})} - 0.08, \qquad
  f_{\mu}=\frac{6.86 \exp(-\lambda / \mu_\mathrm{c}^*)}
  {\lambda^{-1}-\mu_\mathrm{c}^*-{\omega_\mathrm{log} / \bar{\omega}_{2}}}+1.
\end{equation}
In these expressions, $\bar{\omega}_2$ is the square root of the second moment of the normalized weight function
$g(\omega)=2\alpha^2F(\omega)/(\omega\lambda)$, with $\alpha^2F$ being the Eliashberg spectral function, as defined in Ref.~\cite{Giustino2017z}. Equations~\eqref{xsceq1} and \eqref{xsceq2} are useful for preliminary calculations but do not carry the predictive power of the \textit{ab initio} Eliashberg theory.

The Eliashberg theory~\cite{Eliashberg1960s,Eliashberg1961s} describes the superconducting phase transition by means of finite-temperature Green's functions. In this theory, superconducting pairing arises from an attractive electron-electron interaction mediated by phonons, which is partly compensated by the inter-electron Coulomb repulsion. The Eliashberg theory can be formulated as a Dyson equation for a generalized $2\times2$ matrix Green's function via the Nambu-Gor'kov formalism~\cite{Gorkov1958s,Nambu1960s}. The off-diagonal elements of this matrix describe Cooper-pair amplitudes in the superconducting state and are related to the superconducting gap function. These elements become nonzero below the critical temperature, marking the transition to the superconducting state. It is standard practice to expand the pairing self-energy using Pauli matrices $\hat{\tau}_{i}$ ($i= 0,\dots,3$) as follows~\cite{Scalapino1966s,Scalapino1969s,Allen1983s,Carbotte1990s,Choi2003s,Marsiglio2008s,Margine2013s,Marsiglio2020s}:
\begin{align}\label{xsceq12}
\hat{\Sigma}_{n \mathbf{k}} & (i\omega_{j}) = 
i \hbar\omega_{j}\left[1-Z_{n \mathbf{k}}(i \omega_{j})\right] \hat{\tau}_{0} +\chi_{n \mathbf{k}}(i \omega_{j}) \hat{\tau}_{3}
+\phi_{n \mathbf{k}}(i \omega_{j}) \hat{\tau}_{1},
\end{align}
where $i\omega_j = i(2j + 1)\pi T$ is the fermionic Matsubara frequency with $j$ being an integer, $T$ is the absolute temperature, $Z_{n \mathbf{k}}(i \omega_{j})$ is the mass renormalization function, $\chi_{n \mathbf{k}}(i \omega_{j})$ is the energy shift, and $\phi_{n \mathbf{k}}(i \omega_{j})$ is the order parameter. 
This self-energy is expressed in terms of the electron Green's function using the Migdal approximation~\cite{Migdal1958s} for the electron-phonon contribution, and the GW approximation for the electron-electron contribution~\cite{Hedin1965s,Hybertsen1986s}. Using Eq.~\eqref{xsceq12} inside the Dyson equation for the electron Green's function yields a set of coupled equations for $Z_{n \mathbf{k}}$, $\chi_{n \mathbf{k}}$, and $\phi_{n \mathbf{k}}$: \begin{align}
Z_{n \mathbf{k}}&(i \omega_{j})=
1+\frac{\kt}{\omega_{j}N(\varepsilon_{\mathrm{F}})} \sum_{m j^{\prime}} \int\!\frac{d\bq}{\Omega_{\rm BZ}}\,\frac{\omega_{j^{\prime}}
Z_{m \mathbf{k}+\mathbf{q}}(i \omega_{j^{\prime}})}{\theta_{m \mathbf{k}+\mathbf{q}}
(i \omega_{j^{\prime}})}
\lambda(n \mathbf{k}, m \mathbf{k}+\mathbf{q},
\omega_{j}-\omega_{j^{\prime}}),\label{xsceq16}\\
\chi_{n \mathbf{k}}&(i \omega_{j})=-\frac{\kt}{N(\varepsilon_{\mathrm{F}})} \sum_{m j^{\prime}}\int\!\frac{d\bq}{\Omega_{\rm BZ}}\,
\frac{\varepsilon_{m \mathbf{k}+\mathbf{q}}-\mu_{\rm F}+
\chi_{m \mathbf{k}+\mathbf{q}}(i \omega_{j^{\prime}})}{\theta_{m \mathbf{k}+\mathbf{q}}(i \omega_{j^{\prime}})}
\lambda(n \mathbf{k}, m \mathbf{k}+\mathbf{q},
\omega_{j}-\omega_{j^{\prime}}),\label{xsceq17}\\
\phi_{n \mathbf{k}}&(i \omega_{j})=
\frac{\kt}{N(\varepsilon_{\mathrm{F}})} \sum_{m j^{\prime}} \int\!\frac{d\bq}{\Omega_{\rm BZ}}\,\frac{\phi_{m \mathbf{k}+\mathbf{q}}
(i \omega_{j^{\prime}})}{\theta_{m \mathbf{k}+\mathbf{q}}(i \omega_{j^{\prime}})}
\left[\lambda(n \mathbf{k}, m \mathbf{k}+\mathbf{q},
\omega_{j}-\omega_{j^{\prime}})
-N(\varepsilon_{\mathrm{F}})V_{n \mathbf{k}, m \mathbf{k}+\mathbf{q}}\right],\label{xsceq18}
\end{align}
having defined the auxiliary function:
\begin{align}\label{xsceq15}
\theta_{n \mathbf{k}}(i \omega_{j})=&\left[\hbar\omega_{j} Z_{n \mathbf{k}}
(i \omega_{j})\right]^{2} +\left[\varepsilon_{n \mathbf{k}}-
\mu_{\rm F}+\chi_{n \mathbf{k}}(i \omega_{j})\right]^{2}
+\left[\phi_{n \mathbf{k}}(i \omega_{j})\right]^{2}.
\end{align}
In Eqs.~\eqref{xsceq16}-\eqref{xsceq15}, $N(\varepsilon_{\mathrm{F}})$ is the density of states (DOS) per spin at the Fermi level, $\mu_{\rm F}$ is the chemical potential, and the quantities $V_{n \mathbf{k}, m \mathbf{k}+\mathbf{q}}$ denote the matrix elements of the screened Coulomb interaction W between electron pairs, as given in Refs.~\cite{Lee1995,Margine2016s}. The anisotropic electron-phonon coupling parameters $\lambda (n \mathbf{k}, m \mathbf{k}+\mathbf{q}, \omega_{j}-\omega_{j^{\prime}})$ appearing in these equations are calculated as:
\begin{align}\label{xsceq8}
\lambda (n \mathbf{k}, m \mathbf{k}+\mathbf{q}, \omega_{j}-\omega_{j^{\prime}})= 
\frac{N(\varepsilon_{\mathrm{F}})}{\hbar} 
\sum_{\nu}\left|g_{m n \nu}(\mathbf{k}, \mathbf{q})\right|^{2} 
\frac{2 \omega_{\mathbf{q} \nu}}{(\omega_{j}-\omega_{j^{\prime}})^{2}
+\omega_{\mathbf{q} \nu}^{2}}.
\end{align}
Equations~\eqref{xsceq16}-\eqref{xsceq15} are supplemented by a statement of particle number conservation, which determines the chemical potential~\cite{Marsiglio2008s}: 
\begin{equation}\label{xsceq19}
N_{\rm e}=\sum_{n} \int\frac{d\bk}{\Omega_{\rm BZ}} \left[1-2\kt \sum_{j}
\frac{\varepsilon_{n \mathbf{k}}-\mu_{\rm F} + 
\chi_{n \mathbf{k}}(i\omega_j)}{\theta_{n \mathbf{k}}(i\omega_j)}\right],
\end{equation}
where $N_{\rm e}$ is the number of electrons per unit cell. Equations~\eqref{xsceq16}-\eqref{xsceq19} are referred to as the anisotropic full-bandwidth (FBW) Eliashberg equations~\cite{Aperis2018s} since they explicitly take into account scattering processes involving electrons with energies and momenta that are not restricted to the vicinity of the Fermi surface. To find the temperature-dependent superconducting gap, in \texttt{EPW} these equations are solved iteratively for different temperatures; the highest temperature for which a non-trivial solution exists ($\phi_{n\bk}\ne 0$) is the superconducting critical temperature $T_{\mathrm{c}}$. Below this temperature, the superconducting gap $\Delta_{n\bk}$ is given by:
\begin{equation}\label{xsceq20}
\Delta_{n \mathbf{k}}(i \omega_{j})= \phi_{n \mathbf{k}}(i \omega_{j})/Z_{n \mathbf{k}}(i \omega_{j})~.
\end{equation}
The numerical solution of Eqs.~\eqref{xsceq16}-\eqref{xsceq19} is computationally demanding. A common simplification of these equations consists of restricting the energy range close to the Fermi level~\cite{Scalapino1966s,Scalapino1969s,Allen1976s,Allen1983s,Carbotte1990s,Choi2003s,Marsiglio2008s,Margine2013s,Marsiglio2020s}. In this approach, it is assumed that the DOS within this energy window is constant. It can be shown that, within these approximations, the energy shift $\chi_{n\bk}$ vanishes and the requirement in Eq.~\eqref{xsceq19} is automatically satisfied. As a result, only two equations for $Z_{n \mathbf{k}}$ and $\phi_{n \mathbf{k}}$ need  to be solved self-consistently:
\begin{eqnarray}\label{xsceq21}
Z_{n \mathbf{k}}(i \omega_{j})&=& 
1+\frac{\pi \kt}{N(\varepsilon_{\mathrm{F}}) \omega_{j}} \sum_{m 
j^{\prime}} \int\!\frac{d\bq}{\Omega_{\rm BZ}}\,\frac{\omega_{j^{\prime}}}{\sqrt{\hbar^2\omega_{j^{\prime}}^{2}+
\Delta_{m \mathbf{k}+\mathbf{q}}^{2}(i \omega_{j^{\prime}})}} \nonumber \\
 &\times& 
\lambda(n \mathbf{k}, m \mathbf{k}+\mathbf{q}, 
\omega_{j}\!-\!\omega_{j^{\prime}}) \delta(\varepsilon_{m \mathbf{k}+\mathbf{q}}\!-\!
\varepsilon_{\mathrm{F}}),  \\[8pt]
\label{xsceq22}
Z_{n \mathbf{k}}(i \omega_{j}) \Delta_{n \mathbf{k}}(i \omega_{j}) &=&
\frac{\pi \kt}{N(\varepsilon_{\mathrm{F}})}\sum_{m j^{\prime}}\int\!\frac{d\bq}{\Omega_{\rm BZ}}\, 
\frac{\Delta_{m \mathbf{k}+\mathbf{q}}
(i \omega_{j^{\prime}})}{\sqrt{\hbar^2\omega_{j^{\prime}}^{2}+
\Delta_{m \mathbf{k}+\mathbf{q}}^{2}(i \omega_{j^{\prime}})}} \nonumber\\
&\times&\left[\lambda(n \mathbf{k}, m \mathbf{k}+\mathbf{q}, 
\omega_{j}-\omega_{j^{\prime}})-N(\varepsilon_{\mathrm{F}}) V_{n \mathbf{k}, m \mathbf{k}+\mathbf{q}}\right] \delta
(\varepsilon_{m \mathbf{k}+\mathbf{q}}-\varepsilon_{\mathrm{F}}).
\end{eqnarray}
These equations are referred to as the anisotropic Fermi surface restricted (FSR) Eliashberg equations~\cite{Margine2013s,Ponce2016a}. 

To extract physical quantities of interest, such as the tunneling density of states and the heat capacity, the gap function on the real frequency axis is required~\cite{Scalapino1969s,Allen1983s,Carbotte1990s}. In the \texttt{EPW} code, the continuation of $\Delta_{n \mathbf{k}}(i \omega_{j})$ from the imaginary to the real axis is performed either using Pad{\'{e}} approximants~\cite{Vidberg1977s,Leavens1985s}, or using the exact analytic continuation of Ref.~\cite{Marsiglio1988s}, as described in Ref.~\cite{Margine2013s}.

Accounting for the Coulomb repulsion in the Eliashberg equations requires the evaluation of the matrix elements $V_{n \mathbf{k}, m \mathbf{k}+\mathbf{q}}$. While this is feasible in principle, it is common practice to  replace the product $N(\varepsilon_{\mathrm{F}}) V_{n \mathbf{k}, m \mathbf{k}+\mathbf{q}}$ with the semi-empirical Morel-Anderson pseudopotential~$\mu_\mathrm{c}^*$~\cite{Morel1962s}. This is also the approach currently used in \texttt{EPW}, where $\mu_\mathrm{c}^*$ is specified by the user as an external parameter. In many applications, values in the range $\mu_\mathrm{c}^*=0.1$–0.2 yield reasonable agreement with experiments. More recently, first-principles calculations of $\mu_\mathrm{c}^*$ as a Fermi surface average of $V_{n \mathbf{k}, m \mathbf{k}+\mathbf{q}}$ have been used in conjunction with \texttt{EPW}~\cite{Margine2016s,Heil2017s,Heil2019s,DiCataldo2021s}.

In the case of simple superconductors which do not exhibit significant anisotropy, it may be a good approximation to neglect the band and momentum dependence of the superconducting gap. In these cases, instead of Eqs.~\eqref{xsceq16}-\eqref{xsceq19}, it is possible to solve a simplified version of the equations where all properties are averaged over the Fermi surface~\cite{Pickett1982s,Ummarino2013s,Sanna2018s,Davydov2020s}. The relations are referred to as the isotropic Eliashberg equations:
\begin{align}
Z(i \omega_{j}) =\, & 1+\frac{\kt}{N(\varepsilon_{\rm F})\omega_{j}}\int d\varepsilon N(\varepsilon ) \sum_{j^{\prime}} \frac{\omega_{j^{\prime}}
Z(i \omega_{j^{\prime}})} {\theta(\varepsilon,i \omega_{j^{\prime}})}
\lambda(\omega_{j}-\omega_{j^{\prime}}),\label{xsceq23}\\
\chi(i \omega_{j}) =\, &
 -\frac{\kt}{N(\varepsilon_{\rm F})} \int d\varepsilon N(\varepsilon ) \sum_{j^{\prime}} \frac{\varepsilon-\mu_{\rm F}+\chi(i \omega_{j^{\prime}})}
{\theta(\varepsilon, i \omega_{j^{\prime}})}
\lambda(\omega_{j}-\omega_{j^{\prime}}), \label{xsceq24}\\
\phi(i \omega_{j}) =\, &
\frac{\kt}{N(\varepsilon_{\rm F})} \int d\varepsilon N(\varepsilon ) \sum_{j^{\prime}} \frac{\phi(i \omega_{j^{\prime}})}
{\theta(\varepsilon, i \omega_{j^{\prime}})}
\left[\lambda(\omega_{j}-\omega_{j^{\prime}})-\mu^*_c\right],\label{xsceq25}\\
N_{\rm e} =\, &
\int d\varepsilon N(\varepsilon) \left[1 - \frac{2\kt}{N(\varepsilon_{\mathrm{F}})} \sum_{j} \frac{\varepsilon-\mu_{\rm F} + 
\chi(i\omega_j)}{\theta(\varepsilon, i\omega_j)} \right], \label{xsceq29}
\end{align}
where the counterpart of Eq.~\eqref{xsceq15} is:
\begin{equation}\label{xsceq27}
\theta(\varepsilon, i \omega_{j})= 
\left[\hbar\omega_{j}Z(i \omega_{j})\right]^2
+\left[ \varepsilon - \mu_{\rm F} + \chi(i\omega_{j})\right]^2
+\left[ \phi(i \omega_{j})\right]^2.
\end{equation}
In Eqs.~(\ref{xsceq23})-(\ref{xsceq25}), $\lambda(\omega_{j}-\omega_{j^{\prime}})$ is a momentum-averaged version of Eq.~\eqref{xsceq8}:
\begin{align}\label{xsceq28}
\lambda(\omega_{j}-\omega_{j^{\prime}})&= \frac{1}{\left[N(\varepsilon_{\mathrm{F}})\right]^2} 
\sum_{m,n} 
\int\!\frac{d\bk}{\Omega_{\rm BZ}}\,
\int\!\frac{d\bq}{\Omega_{\rm BZ}}\,
\lambda (n \mathbf{k}, m \mathbf{k}+\mathbf{q}, \omega_{j}-\omega_{j^{\prime}})
\delta(\varepsilon_{n \mathbf{k}}-\varepsilon_{\mathrm{F}})
\delta(\varepsilon_{m \mathbf{k}+\mathbf{q}}-\varepsilon_{\mathrm{F}}).
\end{align}
Similar to the anisotropic case, Eqs.~(\ref{xsceq23})-(\ref{xsceq25})  can also be reduced to a set of two equations for $Z(i\omega_{j})$ and $\phi(i \omega_{j})$, under the same conditions as discussed for Eqs.~\eqref{xsceq21}-\eqref{xsceq22}.

\subsubsection{Computational considerations}

Solving Eqs.~\eqref{xsceq16}-\eqref{xsceq19} or the simplified version Eqs.~\eqref{xsceq21}-\eqref{xsceq22} is computationally demanding because a fine sampling of electron-phonon processes near the Fermi surface is required. In addition, due to the implicit form of these equations, the $\mathbf{k}$- and  $\mathbf{q}$-point grids must be commensurate. To accelerate calculations, \texttt{EPW} exploits crystal symmetries so that the gap function and related quantities are only computed for $\mathbf{k}$-points in the irreducible wedge of the Brillouin zone, and only electronic states within a small energy window centered at the Fermi energy are considered. Numerical tests show that converged results are obtained when this energy window is of the order of a few times the maximum phonon energy. 

One further complication in the solution of the Eliashberg equations is that all quantities depend on Matsubara frequencies. The Matsubara frequencies are proportional to the absolute temperature, therefore superconductors with low $T_{\rm c}$ require a larger number of frequencies within the same energy range. While it is common to set the Matsubara frequency cutoff to ten times the maximum phonon frequency~\cite{Pickett1982s,Margine2013s,Sano2016s,Marsiglio2020s,Schrodi2020s}, we recommend to always perform convergence tests with respect to this parameter. In \texttt{EPW}, the number of Matsubara frequencies can be reduced  using a sparse sampling scheme whereby only a subset of frequencies is retained with a density that decreases logarithmically with the Matsubara index. The grid density is controlled by an adjustable parameter; using the default setting for this parameter, \texttt{EPW} yields a $\sim$30\% reduction of Matsubara frequencies, while keeping all the lowest $\sim$40\% of frequencies. Numerical tests show that this approach preserves the accuracy of more expensive full-grid calculations.

FBW Eliashberg calculations are computationally more demanding than FSR calculations. In particular, to determine the chemical potential from Eq.~\eqref{xsceq19} one needs a Matsubara frequency cutoff of at least a few times the Kohn-Sham energy window, leading to a considerable increase in computational cost. To circumvent this issue, the default behavior of \texttt{EPW} is to keep the chemical potential fixed at the Fermi level, unless it is instructed by the user to solve Eq.~(\ref{xsceq19}). In this case, careful convergence tests with respect to the  Matsubara frequency cutoff are warranted.

\subsubsection{Application example}\label{scexample}

To demonstrate the implementation of the superconducting module in \texttt{EPW}, we investigate the superconducting properties of hexagonal niobium diselenide (2H-NbSe$_2$). 2H-NbSe$_2$ exhibits a superconducting phase transition and a charge density wave instability below 7.2~K and 33~K, respectively; both phases are driven by a strong momentum-dependent electron-phonon coupling~\cite{Valla2004s,Weber2011s,Leroux2015s,Yokoya2001s,Anikin2020s,Sanna2022s}. For brevity, in this section we focus on superconductivity and ignore the lattice instability.

We employ the PBE exchange and correlation functional~\cite{Perdew1996p} and the optB86b-vdW van der Waals functional~\cite{Klimes2011s,Klimes2010s,Thonhauser2015s,Thonhauser2007s}, as well as ONCV pseudopotentials~\cite{Hamann2013s,Schlipf2015s} including semi-core electrons for Nb. We use a planewaves 
kinetic energy cutoff of 80~Ry, $\Gamma$-centered 24$\times$24$\times$12 $\mathbf{k}$-point and  6$\times$6$\times $4 $\mathbf{q}$-point coarse grids, and Methfessel-Paxton smearing~\cite{Methfessel1989s} of 0.025~Ry. The optimized lattice parameters are $a=3.46$ \AA~and $c=12.57$ \AA, in good agreement with the experimental data $a=3.43$ \AA~and $c=12.55$ \AA~\cite{Weber2011s}. 

The Eliashberg equations are solved on uniform 60$\times$60$\times$40 $\mathbf{k}$-point and 30$\times$30$\times $20 $\mathbf{q}$-point fine grids. We consider energy windows of $0.4$ and $0.8$~eV, Matsubara frequency cutoffs of 0.4, 0.8, and 1.2~eV, and a Coulomb pseudopotential of $\mu_\mathrm{c}^* = 0.2$. The Dirac deltas in the FSR approach are replaced by Gaussians of width $50$~meV.

\begin{figure}
	\centering
    \includegraphics[width=\textwidth]{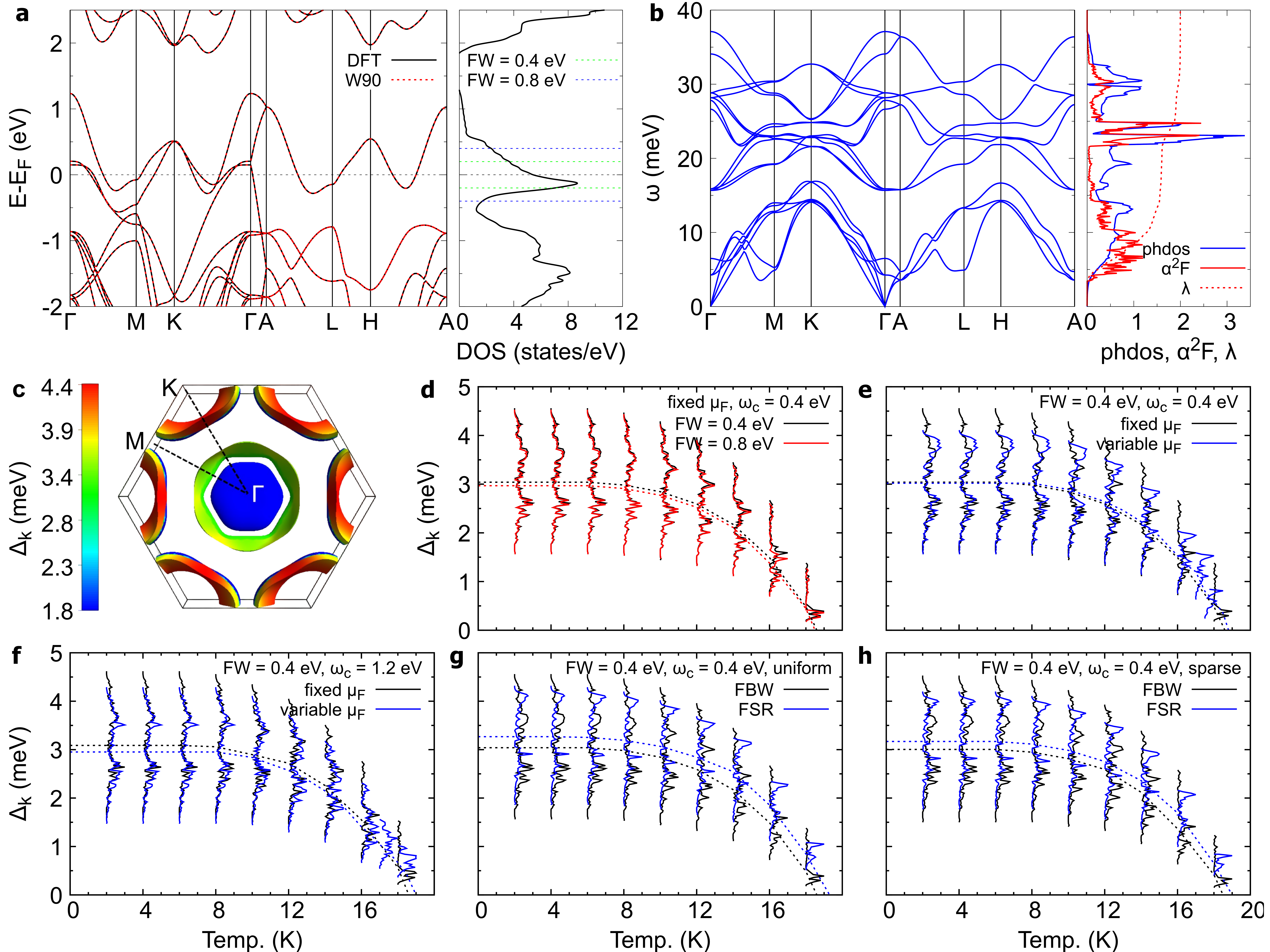}
	\caption{\textbf{Eliashberg calculations for phonon-mediated superconductors in \texttt{EPW}.} (a) Band structure and DOS of 2H-NbSe$_2$. (b) Phonon dispersion relations, phonon density of states (blue line), Eliashberg spectral function $\alpha^2F$ (solid red line), and cumulative electron-phonon coupling strength $\lambda$ (dashed red line) of 2H-NbSe$_2$. In (a), the bands obtained by Wannier interpolation are shown in dashed red lines. (c) Momentum-resolved superconducting gap of 2H-NbSe$_2$, color-coded on the Fermi surface~\cite{Kawamura2019s} and evaluated at 2~K [same dataset as in (d)]. (d) Energy distribution of the superconducting gap of 2H-NbSe$_2$ as a function of temperature, calculated using the FBW Eliashberg approach, chemical potential set to the Fermi energy, and a Matsubara frequency cutoff of 0.4~eV. The energy window is 0.4~eV for the black line, and 0.8~eV for the red line. (e) Gap function of 2H-NbSe$_2$ obtained from the FBW approach using fixed chemical potential (black line) and variable chemical potential (blue line). The energy window and Matsubara cutoff are both 0.4~eV. (f) Same as in (e), but using a Matsubara cutoff of 1.2~eV. (g) Comparison between the gap functions obtained with the FBW approach (black line) and the FSR approach (blue). Both the energy window and the Matsubara cutoff are 0.4~eV. (h) Same as in (g), but using a sparse Matsubara frequency grid. The dashed lines in (d)-(h) represent averages of the gap distributions.}
	\label{fig-sc} 
\end{figure}

Figure~\ref{fig-sc}(a) shows the band structure and DOS of 2H-NbSe$_2$. Three bands cross the Fermi level. This system exhibits soft phonons along the $\Gamma M$ direction~\cite{Leroux2015s,Sanna2022s,Anikin2020s}, which are stabilized by anharmonic effects~\cite{Leroux2015s}. Here, for simplicity, we avoid soft phonons by slightly increasing the electronic smearing to 0.025~Ry in phonon calculations, as shown in Fig.~\ref{fig-sc}(b). The same panel also shows the distribution of the electron-phonon coupling strength as described by the Eliashberg spectral function $\alpha^2F$.

Figure~\ref{fig-sc}(c) shows the anisotropic gap function color-coded on the Fermi surface of 2H-NbSe$_2$. We find a highly anisotropic two-gap structure: the high-gap region corresponds to the Fermi arcs around the $\rm K$ point, and the low-gap region corresponds to the $\Gamma$-centered pockets. These results are similar to previous calculations for the related compound NbS$_2$~\cite{Heil2017s}.

Figure~\ref{fig-sc}(d) shows the energy distribution of the superconducting gap function $\Delta_{n\bk}$ for various temperatures. The gap is seen to close gradually with increasing temperature, vanishing at the critical temperature $T_{\rm c}=19$~K. The calculated critical temperature overestimates the experimental value of 7.2~K, consistent with prior calculations for bulk and monolayer transition metal dichalcogenides~\cite{Heil2017s,Leroux2015s,Sanna2022s,Zheng2019s,Wickramaratne2020s,Das2022s}. The calculations in Fig.~\ref{fig-sc}(d) are performed within the FBW Eliashberg approach, with the chemical potential set to the Fermi energy. Calculated gap values for energy windows of $0.4$~eV and $0.8$~eV are very similar, therefore in the following we only consider the former.

Figure~\ref{fig-sc}(e) shows the sensitivity of the superconducting gap to the chemical potential: by requiring particle number conservation via Eq.~\eqref{xsceq19}, the chemical potential shifts by 25~meV with respect to the Fermi energy, and the critical temperature slightly increases, by less than 1~K. In Fig.~\ref{fig-sc}(f) we perform the same test, but this time with a wider energy window, obtaining similar results.

In Fig.~\ref{fig-sc}(g) we also compare the FBW Eliashberg approach with the FSR approach, which is approximately 30\% faster for the system under consideration. The critical temperatures calculated from these approaches agree within 0.5~K. This finding relates to the DOS of 2H-NbSe$_2$ being slowly varying within a few hundred meV's around the Fermi energy. Larger differences are expected for systems with rapidly varying DOS, such as for example the high-$T_{\rm c}$ hydride superconductors~\cite{Sano2016s}.

When using a logarithmic Matsubara frequency grid, the results are essentially unaffected as shown in Fig.~\ref{fig-sc}(h). At the same time, the sparse Matsubara grids afford a significant computational saving of almost a factor of two.\\

  \subsection{Polarons and electron self-trapping}\label{sec.polarons}

\subsubsection{Background and formalism}

An important manifestation of electron-phonon interactions is the formation of polarons. The polaron is a quasiparticle consisting of an electron or a hole dressed by a distortion of the crystal lattice. The lattice distortion can be interpreted as a cloud of virtual phonons accompanying the electron or hole. In the presence of strong electron-phonon coupling, the electron or hole can become trapped in the lattice distortion that it produced; this configuration is referred to as a self-trapped polaron~\cite{Alexandrov2010p,Emin2012p,Franchini2021p}.

In the context of first-principles calculations, the formation of polarons has generally been investigated by means of DFT calculations of an excess charge added to a large supercell. When the calculation is initialized with a localized distortion of the crystal lattice, it may be energetically favorable for the electron or hole to localize around this distortion, leading to the formation of a polaron~\cite{Deskins2007p,Franchini2009p,Lany2009p,Varley2012p,Setvin2014p,Kokott2018p,Falletta2022p}.
This approach faces two challenges: first, the existence of polaronic states and their formation energy are very sensitive to the choice of the exchange and correlation functional; second, the size of the supercells required to obtain converged energies and wavefunctions may be prohibitive. 

In the \texttt{EPW} code, these challenges are overcome by recasting the polaron problem into the solution of a coupled nonlinear system of equations for the energy, wavefunction, and atomic displacements associated with the polaron~\cite{Sio2019pa,Sio2019pb,Lafuente2022pa,Lafuente2022pb}. The founding principle of this methodology is that the DFT formation energy of the polaron can be expressed as a self-interaction-free functional of the polaron wavefunction $\psi(\br)$ and the atomic displacements in the polaronic state $\Delta\tau_{\kappa\alpha p}$~\cite{Sio2019pb}:  
\begin{equation}\label{eq:pol-funct}
    \Delta E_{\rm f}
    =
    \int\!\! d\mathbf{r} \, \psi^{*}(\mathbf{r}) \hat{H}_{\mathrm{KS}}^{0} \psi(\mathbf{r})
    + \sum_{\kappa\alpha p}\int\!\! d\mathbf{r} \, \frac{\partial V^{0}_{\mathrm{KS}}}{\partial \tau_{\kappa\alpha p}}
    |\psi(\mathbf{r})|^{2} \Delta\tau_{\kappa\alpha p}
    + \frac{1}{2} \sum_{\substack{\kappa\alpha p\\\kappa'\alpha' p'}} C^{0}_{\kappa\alpha p, \kappa'\alpha' p'} \Delta\tau_{\kappa\alpha p} \Delta\tau_{\kappa'\alpha' p'},
\end{equation}
where $\hat{H}_{\mathrm{KS}}^{0}$ and $C^{0}_{\kappa\alpha p, \kappa'\alpha' p'}$ are the Kohn-Sham Hamiltonian and the matrix of interatomic force constants in the ground-state structure without the polaron, and the integrals are over the BvK supercell. Variational minimization of this energy functional leads to the coupled system of equations:
\begin{eqnarray}
    &&\hat{H}_{\mathrm{KS}}^{0} \, \psi(\mathbf{r})
    +
    \sum_{\kappa\alpha p} \frac{\partial V_{\mathrm{KS}}^{0}(\mathbf{r})}{\partial \tau_{\kappa \alpha p}}
    \Delta \tau_{\kappa \alpha p} \, \psi(\mathbf{r})
    =
    \varepsilon \, \psi(\mathbf{r}) ~, \label{eq:psir} \\
    &&
    \Delta \tau_{\kappa \alpha p}
    =
    -\sum_{\kappa'\alpha' p'} (C^{0})^{-1}_{\kappa\alpha p,\kappa'\alpha'p'}
    \!\int\!\! d\mathbf{r} \, \frac{\partial V_{\mathrm{KS}(\mathbf{r})}^{0}}{\partial \tau_{\kappa'\alpha'p'}}
    \, |\psi(\mathbf{r})|^2 ~,\label{eq:dtau}
\end{eqnarray}
where $V_{\mathrm{KS}}^{0}$ is the Kohn-Sham potential in the ground-state structure without polaron, and $\varepsilon$ represents the quasiparticle excitation energy of the polaron~\cite{Lafuente2022pa,Lafuente2022pb}. To avoid performing calculations in large supercells, in \texttt{EPW} Eqs.~\eqref{eq:psir} and \eqref{eq:dtau} are rewritten more conveniently in terms of Kohn-Sham states, phonons, and electron-phonon matrix elements evaluated in the primitive unit cell of the crystal. To this aim, the wavefunction is expanded in the basis of Kohn-Sham states:
\begin{equation} \label{eq:exp1}
  \psi(\mathbf{r}) = \frac{1}{\sqrt{\Np}}
  \sum_{n\mathbf{k}} A_{n\mathbf{k}} \psi_{n\mathbf{k}} ~,
\end{equation}
and the atomic displacements are expanded in the basis of lattice vibrational eigenmodes:
\begin{equation} \label{eq:exp2}
  \Delta\tau_{\kappa\alpha p } = -\frac{2}{\Np}
  \sum_{\mathbf{q}\nu} B^{*}_{\mathbf{q}\nu} 
  \sqrt{\frac{\hbar}{2M_\kappa \omega_{\mathbf{q}\nu}}}\,
  e_{\kappa\alpha,\nu}(\mathbf{q}) e^{i\mathbf{q}\cdot\mathbf{R}_p}.
\end{equation}
With these definitions, Eqs.~\eqref{eq:psir} and \eqref{eq:dtau} become a nonlinear system of equations for the polaron coefficients $A_{n\bk}$ and $B_{\bq\nu}$:
\begin{eqnarray} \label{eq:ank}
  &&\frac{2}{\Np} \sum_{\mathbf{q}m\nu} B_{\mathbf{q}\nu}
  \, g_{mn\nu}^{*}(\mathbf{k},\mathbf{q}) \, A_{m\mathbf{k+q}}
  =
  (\varepsilon_{n\mathbf{k}}-\varepsilon) A_{n\mathbf{k}} ~,\\
  && \label{eq:bqv}
  B_{\mathbf{q}\nu} = \frac{1}{\Np}
  \sum_{mn\mathbf{k}} A^{*}_{m\mathbf{k+q}}
  \frac{g_{mn\nu}(\mathbf{k},\mathbf{q})}{\hbar\omega_{\mathbf{q}\nu}} A_{n\mathbf{k}} ~.
\end{eqnarray}
These relations are referred to as the \textit{ab initio} polaron equations.
The polaron formation energy in Eq.~\eqref{eq:pol-funct} can be expressed in terms of the polaron coefficients $A_{n\bk}$ and $B_{\bq\nu}$ as~\cite{Sio2019pb}:
\begin{equation} \label{eq:eform}
  \Delta E_{\rm f}
  =
  \frac{1}{\Np} \sum_{n\mathbf{k}} |A_{n\mathbf{k}}|^2
  (\varepsilon_{n\mathbf{k}}-\varepsilon_{\mathrm{CBM}})
  -
  \frac{1}{\Np} \sum_{\mathbf{q}\nu} |B_{\mathbf{q}\nu}|^2 \hbar\omega_{\mathbf{q}\nu} ~,
\end{equation}
where $\varepsilon_{\mathrm{CBM}}$ is the Kohn-Sham eigenvalue of the conduction band minimum. This expression holds for electron polarons; in the case of hole polarons, $\varepsilon_{\mathrm{CBM}}$ is replaced by the eigenvalue of the valence band maximum, and the first term in the righ-hand side of Eq.~\eqref{eq:eform} acquires a minus sign.
Equations \eqref{eq:exp1}, \eqref{eq:exp2}, and \eqref{eq:eform} provide the real-space wavefunction of the polaron, the accompanying atomic displacements, and its formation energy.

\subsubsection{Computational considerations}

The solution of Eqs.~\eqref{eq:ank} and \eqref{eq:bqv} requires the knowledge of the Kohn-Sham energies $\varepsilon_{n\mathbf{k}}$, the phonon frequencies $\omega_{\mathbf{q}\nu}$, and the electron-phonon matrix elements $g_{mn\nu}(\mathbf{k},\mathbf{q})$. All of these quantities are calculated by \texttt{EPW} using Wannier-Fourier interpolation as described in Sec.~\ref{sec.epwtheory}. The solution of the coupled system of equations is performed using an iterative procedure as follows. First, $A_{n\bk}$ is initialized in order to compute $B_{\bq\nu}$ via Eq.~\eqref{eq:bqv}. Then, Eq.~\eqref{eq:ank} is solved by constructing and diagonalizing the effective Hamiltonian
\begin{equation}\label{eq:bqu2h}
H_{n\bk,n'\bk'}=\varepsilon_{n\mathbf{k}}\delta_{n\bk,n'\bk'}-\frac{2}{\Np} {\sum}_{\nu} B_{\mathbf{k'-k}\nu}  \, g_{n'n\nu}^{*}(\mathbf{k},\mathbf{k'-k}) ~.
\end{equation}
The new set of solution coefficients $A_{n\bk}$ is used again in Eq.~\eqref{eq:bqv}, and the process is repeated until convergence is achieved.

The initialization of the coefficients $A_{n\bk}$ is achieved by using a Gaussian function in reciprocal space centered around the band extremum. Alternatively, a polaron solution from a previous calculation can be used. In the construction of the effective Hamiltonian matrix in Eq.~\eqref{eq:bqu2h}, we set the gauge condition $e_{\kappa\alpha,\nu}(-\bq) = e^*_{\kappa\alpha,\nu}(\bq)$~\cite{Maradudin1968p}. This condition ensures that $B_{-\bq\nu} = B^*_{\bq\nu}$, so only half of the $B_{\bq\nu}$ coefficients need to be evaluated at each iteration. In addition, for parallel execution we set a global gauge for all interpolated Kohn-Sham wavefunctions and vibrational eigenmodes.
The diagonalization of the effective Hamiltonian is performed via the Davidson method~\cite{Davidson1975p} as implemented in \texttt{Quantum ESPRESSO}. The convergence of the iterative procedure is tested by evaluating the atomic displacements in real space via Eq.~\eqref{eq:exp2}, and comparing these real-valued vectors between successive iterations. Since these calculations describe an excess charge in a periodic BvK supercell, the resulting energy needs to be extrapolated to the limit of infinite supercell size; this is achieved by means of the standard Makov-Payne method~\cite{Makov1995p,Sio2019pb}.

To visualize the polaron, \texttt{EPW} expresses the wavefunction as a linear combination of maximally-localized Wannier functions:
\begin{equation}
  \label{eq:amp2psir}
  \psi(\br) = \sum_{mp} A_{mp} {\rm w}_m(\br-\bR_p),
\end{equation}
where the coefficients $A_{mp}$ are obtained from the solution vectors $A_{n\bk}$ via the Wannier-Fourier transformation:
\begin{equation}
\label{eq:ank2amp}
A_{mp} =  \frac{1}{\Np} \sum_{n\bk}e^{i\bk\cdot \bR_p}U^\dagger_{mn\bk} A_{n\bk} ~,
\end{equation}
and $U_{mn\bk}$ is the unitary matrix that generates the smooth Bloch gauge, Eq.~\eqref{eq.wannier-def}. The inversion of this last relation also allows one to interpolate the coefficients $A_{n\bk}$ throughout the Brillouin zone, which is useful to visualize how specific bands contribute to the polaron wavefunction (Fig.~\ref{fig:polaron2}). 

\subsubsection{Application example}

To demonstrate the implementation of the polaron module in \texttt{EPW}, we investigate the electron and hole polarons in rocksalt NaCl, a prototypical polar insulator that hosts small hole polarons called V$_{\rm K}$ centers~\cite{Castner1957p, Norman1969p}. We perform calculations using the PBE exchange and correlation functional~\cite{Perdew1996p}, ONCV pseudopotentials~\cite{setten2018p}, and a planewaves kinetic energy cutoff of 150~Ry. Ground-state and lattice-dynamical calculations are performed using a 12$\times$12$\times$12 uniform and unshifted Brillouin zone mesh for both $\bk$- and $\bq$-points. The Kohn-Sham states, phonons, and electron-phonon matrix elements needed to solve Eqs.~\eqref{eq:ank} and \eqref{eq:bqv} are generated on Brillouin zone grids with up to 50$\times$50$\times$50 points 
by Wannier-Fourier interpolation. To this end, we use three Wannier functions to describe the Cl-$3p$ states in the valence band, and one Wannier function to describe a single conduction band formed by the Na-$3s$ states. The use of additional valence or conduction bands only brings negligible changes to the polaron formation energy. In order to obtain the polaron formation energies in the limit of infinite supercell size, we solve Eqs.~\eqref{eq:ank} and \eqref{eq:bqv} for $N\!\times\!N\times\!N$~$\bk$- and $\bq$-point grids with increasing $N$, and we perform a linear extrapolation of the energy vs.\ $1/N$ curve. 

\begin{figure}
\includegraphics[width=1.0\linewidth]{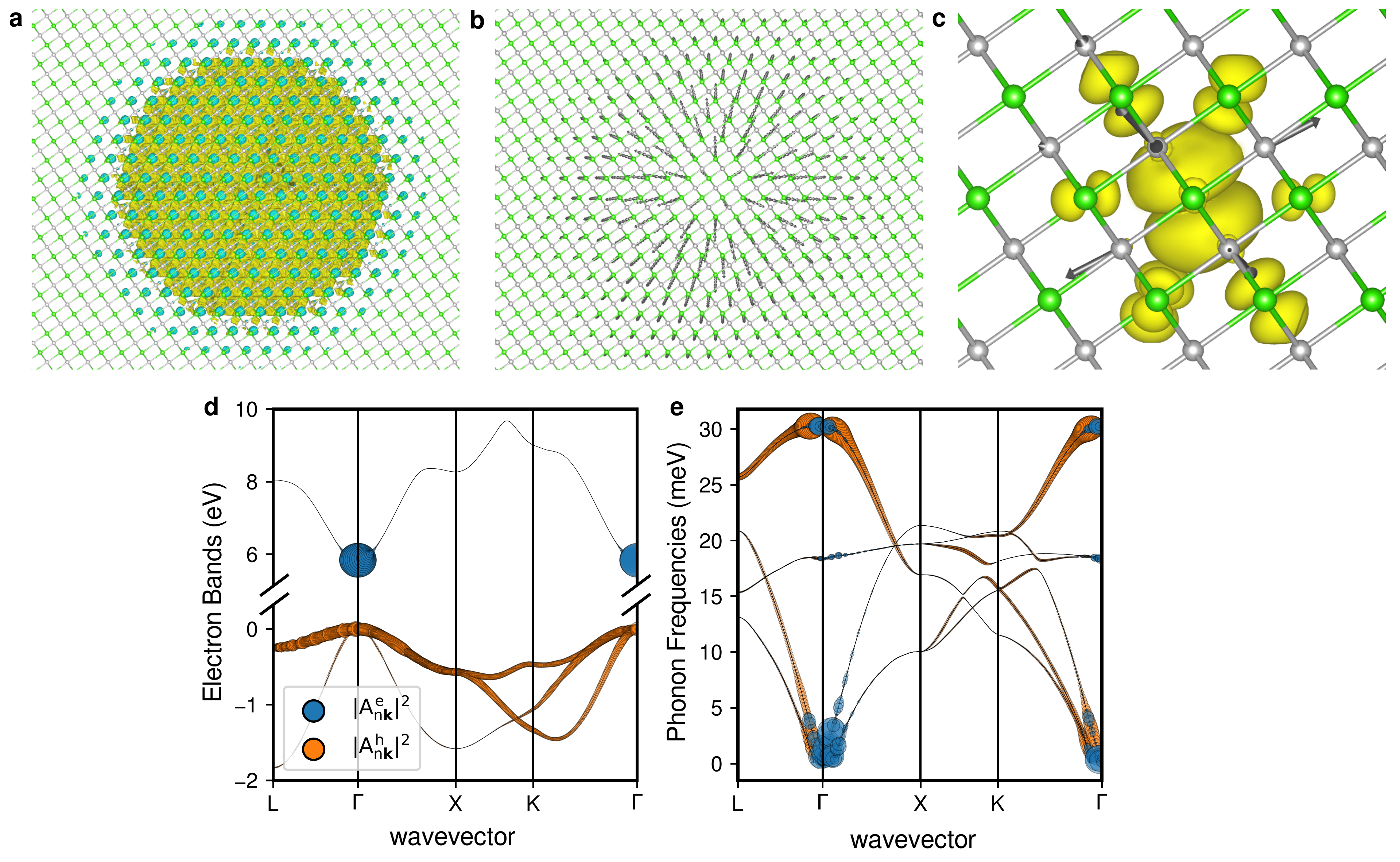}
\caption{\label{fig:polaron2} 
\textbf{Calculation of small and large polarons with \texttt{EPW}.} 
(a) Isosurface plot of the wavefunction of the electron polaron in NaCl. Na atoms are in silver, Cl atoms in green.
(b) Displacements of the Cl atoms associated with the electron polaron shown in (a). The displacements are exaggerated for clarity.
(c) Isosurface plot of the hole polaron in NaCl, and associated atomic displacements. The displacements are exaggerated for clarity.
(d) Spectral weights of the electron (blue) and hole (orange) polaron of NaCl, superimposed to the band structure. The size of the circles is proportional to $|A_{n\bk}|^2$.
(e) Spectral weights of the electron (blue) and hole (orange) polaron of NaCl, superimposed to the phonon dispersion relations. The size of the circles is proportional to $|B_{\bq\nu}|^2$.
}
\end{figure}

Figure~\ref{fig:polaron2}(a) shows an isosurface of the calculated wavefunction of the electron polaron in NaCl. The formation energy of this polaron in the infinite supercell limit is 37~meV; with a radius of 147~\AA, this polaron spans several tens of crystalline unit cells. The wavefunction is isotropic and can be described as a combination of Na-$3s$ orbitals modulated by an approximately Gaussian envelope.

In Fig.~\ref{fig:polaron2}(b) we show the atomic displacements associated with the electron polaron of Fig.~\ref{fig:polaron2}(a). For clarity we only show the displacement of the Cl ions. The ions tend to move away from the center of the electron wavefunction, consistent with the fact that the electron polaron tends to repel anions. 

Figure~\ref{fig:polaron2}(c) shows an isosurface of the calculated hole polaron of NaCl. The formation energy in the dilute limit is 820~meV, indicating a strongly bound polaron. Consistent with the large formation energy, we find a very small polaron of radius 6~\AA, which is comparable to the lattice parameter of 5.69~\AA. This polaron consists primarily of a single $3p$ orbital centered on Cl and oriented along the [100] direction. The accompanying atomic displacements are shown in the same panel. As expected from the negative charge of the polaron, the largest displacements are found for the Na cations which tend to move away from the polaron center.

In Fig.~\ref{fig:polaron2}(d) we overlay the coefficients $|A_{n\bk}|^2$ with the band structure in order to determine which electronic states contribute to the electron (blue) and hole (orange) polaron. In the case of the electron polaron, only electrons near the bottom of the conduction band contribute; the narrow distribution of $A_{n\bk}$ near $\bk=0$ is consistent with the large spatial extent of the electron polaron in real space. Conversely, the hole polaron draws weight from the entire Brillouin zone, in line with the strong localization in real space. Similarly, in Fig.~\ref{fig:polaron2}(e) we show the coefficients $|B_{\bq\nu}|^2$ on the phonon dispersion relations. The electron polaron is primarily driven by long-wavelength longitudinal-optical phonons and longitudinal acoustic phonons, while the hole polaron is driven mostly by short-range optical modes. These spectral weight distributions suggest that the electron polaron in NaCl should be classified as a Fr\"{o}hlich-type~\cite{Frohlich1950p} polaron, while the hole polaron should be classified as a Holstein-type polaron~\cite{Holstein1959p}.

  \subsection{Phonon-assisted indirect absorption}\label{sec.indabs}

\subsubsection{Background and formalism}
\label{sec:indabs_theo}

The \texttt{EPW} code has the capability to compute, from first principles, optical absorption spectra in indirect band gap semiconductors by including phonon-assisted optical transitions within second-order time-dependent perturbation theory. Phonon-assisted transitions involve two virtual processes, namely the absorption of a photon, and the absorption or emission of a phonon, in either order. In this section, we outline the general formalism to describe these processes~\cite{bassani1975o,noffsinger2012o}.

We consider a linearly polarized electromagnetic wave with vector potential of amplitude $A_0$, frequency $\omega$, and polarization vector $\mathbf{e}$. In the following equations, the long-wavelength limit and the corresponding electric dipole approximation are understood. Second-order time-dependent perturbation theory states that the transition rate of an electron from an initial Kohn-Sham state $n\bk$ to a final state $m\bk+\bq$ involving a photon and a phonon $\bq\nu$ is~\cite{bassani1975o}:
\begin{eqnarray}
\label{eq:tdpt}
    W_{mn\nu}(\mathbf{k,q};\omega)&=&\nonumber
    \frac{2\pi}{\hbar}
    e^2 A_0^2\sum_{\beta=\pm 1}\left|\mathbf{e}\cdot[\mathbf{S}_{1,mn\nu}(\mathbf{k,q})+\mathbf{S}_{2,mn\nu\beta}(\mathbf{k,q})]\right|^2 \\ & \times&\delta(\varepsilon_{m\mathbf{k+q}}-\varepsilon_{n\mathbf{k}}-\hbar\omega+\beta\hbar\omega_{\mathbf{q}\nu}).
\end{eqnarray}
In this equation, $\beta=+1$ and $-1$ represent phonon emission and absorption processes, respectively. 
$\mathbf{S}_{1,mn\nu}(\mathbf{k,q})$ and $\mathbf{S}_{2,mn\nu\beta}(\mathbf{k,q})$ are the transition amplitudes for the processes illustrated in Fig. \ref{fig:optics}(a): $\mathbf{S}_{1,mn\nu}(\mathbf{k,q})$ refers to a process whereby the electron absorbs a photon, and then absorbs or emits a phonon; $\mathbf{S}_{2,mn\nu\beta}(\mathbf{k,q})$ describes a process whereby phonon absorption/emission takes place, followed by the absorption of a photon. Explicit expressions for these amplitudes are:
\begin{eqnarray}
   \mathbf{S}_{1,mn\nu}(\mathbf{k,\mathbf{q}})&=&\sum_j \frac{g_{mj\nu}(\mathbf{k,q})\textbf{v}_{jn}(\mathbf{k})}{\varepsilon_{j\mathbf{k}}-\varepsilon_{n\mathbf{k}}-\hbar\omega+i\eta}, \label{eq:s1} \\
   \label{eq:s2}
   \mathbf{S}_{2,mn\nu\beta}(\mathbf{k,\mathbf{q}})&=&\sum_j \frac{\textbf{v}_{mj}(\mathbf{k+q})g_{jn\nu}(\mathbf{k,q})}{\varepsilon_{j\mathbf{k+q}}-\varepsilon_{n\mathbf{k}}+\beta\hbar\omega_{\mathbf{q}\nu}+i\eta},
\end{eqnarray}
where $\textbf{v}_{mn}$ denotes velocity matrix elements between the Kohn-Sham states, and the sum extends to all possible occupied and unoccupied states. The energy $\eta$ is a small parameter to avoid singular denominators.
From the transition rates $W_{mn\nu}(\mathbf{k,q};\omega)$, we obtain the imaginary part of the dielectric function by summing over all possible transitions in the Brillouin zone and in the band manifold~\cite{bassani1975o,noffsinger2012o}:
\begin{eqnarray}
\label{eqn:eps2}
    \text{Im}[\epsilon(\omega)]&=&2\frac{\pi e^2}{\epsilon_0 \Omega}\frac{1}{\omega^2}
    \sum_{mn\nu,\beta=\pm1}\int\!\frac{d\bk}{\Omega_{\rm BZ}}\!\int\!\frac{d\bq}{\Omega_{\rm BZ}}\,\Big|
    \mathbf{e}\cdot[\mathbf{S}_{1,mn\nu}(\mathbf{k,q})+\mathbf{S}_{2,mn\nu\beta}(\mathbf{k,q})]\Big|^2 \nonumber \\ &\times& P_{mn\nu\beta}(\mathbf{k,q})\delta(\varepsilon_{m\mathbf{k+q}}-\varepsilon_{n\mathbf{k}}-\hbar\omega+\beta\hbar\omega_{\mathbf{q}\nu}),
\end{eqnarray} 
where the factor of two results from the electron spin in the case of spin-unpolarized systems; this factor is omitted in the case of calculations including spin-orbit coupling. The quantities $P_{mn\nu\beta}(\mathbf{k,q})$ contain the temperature-dependent Fermi-Dirac and Bose-Einstein distributions as follows:
\begin{equation}
\label{eqn:pa}
P_{mn\nu\beta}(\mathbf{k,q})=\left(n_{\mathbf{\mathbf{q}\nu}}+\frac{1+\beta}{2}\right)
f_{n\mathbf{k}}(1-f_{m\mathbf{k+q}})-\left(n_{\mathbf{\mathbf{q}\nu}}+\frac{1-
\beta}{2}\right) (1-f_{n\mathbf{k}})f_{m\mathbf{k+q}}
\end{equation}
In Eq.~\eqref{eqn:eps2}, the scalar quantity $\epsilon(\omega)$ represents the projection of the dielectric tensor along the polarization direction $\mathbf{e}$ of the electric field. For easier comparison with experimental measurements, after we obtain the imaginary part of the dielectric function from \texttt{EPW}, we calculate the absorption coefficient using~\cite{bassani1975o}: 
\begin{equation}
\label{eq:abs}
    \alpha(\omega)=\frac{\omega \,\textrm{Im}[\epsilon(\omega)]}{c\, n(\omega)},
\end{equation}
where $n(\omega)$ is the real part of the refractive index and $c$ is the speed of light. $n(\omega)$ can be calculated from the standard relations between the dielectric function and the refractive index~\cite{giustino2014o}. In this case, the real part of the dielectric function can be obtained by performing a Kramers-Kronig transformation of Im$[\epsilon(\omega)]$, and applying a rigid shift to match the value of $\epsilon(\omega=0)$ computed from DFPT using \texttt{Quantum ESPRESSO}. Alternatively, $n(\omega)$ can be taken from experiments~\cite{noffsinger2012o}.

\subsubsection{Computational considerations}
In the optics module of \texttt{EPW}, the imaginary part of the dielectric function is calculated using Eq.~\eqref{eqn:eps2}.
The summation over virtual states in Eqs.~\eqref{eq:s1} and \eqref{eq:s2} are restricted to the manifold of valence and conduction bands included in the Wannierization procedure. The Dirac delta functions appearing in Eq.~(\ref{eqn:eps2}) are replaced by Gaussian functions or Lorentzian functions with a finite broadening. 
The small parameter $\eta$ in Eqs.~\eqref{eq:s1} and \eqref{eq:s2} is used to avoid singular denominators which arise if, for a given photon energy $\hbar\omega$, direct transitions are resonant with indirect transitions. These situations are encountered, for example, at the onset of direct transitions, i.e. when the photon energy matches the direct gap. In these situations, the dielectric function and the absorption spectrum become sensitive to the choice of $\eta$: too small an $\eta$ leads to a divergence of the spectrum, and too large an $\eta$ leads to an excessive broadening. To probe the sensitivity of the spectra to this parameter, the optics module of \texttt{EPW} calculates $\text{Im}[\epsilon(\omega)]$ for a range of broadening parameters between 1~meV and 0.5~eV.
Efforts are currently ongoing to eliminate these spurious singularities: we believe that they arise from an intrinsic limitation of second-order perturbation theory when direct and indirect transitions are in resonance; a more general theory that correctly describes these resonances is under development~\cite{Tiwari2023}.

In Eqs.~(\ref{eq:s1}) and (\ref{eq:s2}), the velocity matrix elements are sensitive to the details of the electronic structure. For accurate calculations, it is preferable to include GW quasiparticle corrections to the Kohn-Sham eigenvalues~\cite{Hybertsen1986s}. These corrections require a renormalization of the velocity matrix elements to preserve the $f$-sum rule of optical transitions~\cite{Levine1991o,rohlfing2000o}.
In the \texttt{EPW} code, this renormalization is carried out by evaluating the velocity via finite differences. To this end, we consider quasiparticle energies and Kohn-Sham eigenvalues evaluated on $\mathbf{k}$-point grids slightly offset along the Cartesian directions. The renormalized velocity matrix elements are given by\cite{rohlfing2000o}:
\begin{equation}
\label{eq:ren_vme}
        v_{mn,\alpha}^{\rm QP}(\mathbf{k})=\frac{\varepsilon_{m\mathbf{k+\delta q_\alpha}}^{\rm QP}-\varepsilon_{n\mathbf{k-\delta q}_\alpha}^{\rm QP}}{\varepsilon_{m\mathbf{k+\delta q}_\alpha}^{\rm KS}-\varepsilon_{n\mathbf{k-\delta q}_\alpha}^{KS}} v_{mn,\alpha}^{\rm KS}(\mathbf{k}),
\end{equation}
where $\delta\mathbf{q}_\alpha$ is a small wavevector along the Cartesian direction $\alpha$, of magnitude $10^{-3}\times 2\pi/a$ ($a$ is the lattice parameter).

\subsubsection{Application example}

To demonstrate the implementation of the optics module in \texttt{EPW}, we examine phonon-assisted optical absorption in silicon. Gound-state DFT calculations are performed using the PBE exchange and correlation functional~\cite{Perdew1996p}, ONCV pseudopotentials~\cite{Hamann2013s,Schlipf2015s}, a planewaves kinetic energy cutoff of 60~Ry, and a 6$\times$6$\times$6 Brillouin zone sampling of both the $\mathbf{k}$-point grid and the $\mathbf{q}$-point grid. The optimized DFT lattice parameter is $a=5.478$~\AA, similar to previous studies~\cite{mo2018o,haas2009o,favot1999o}.

\begin{figure}
    \centering
    \includegraphics[width=0.9\textwidth]{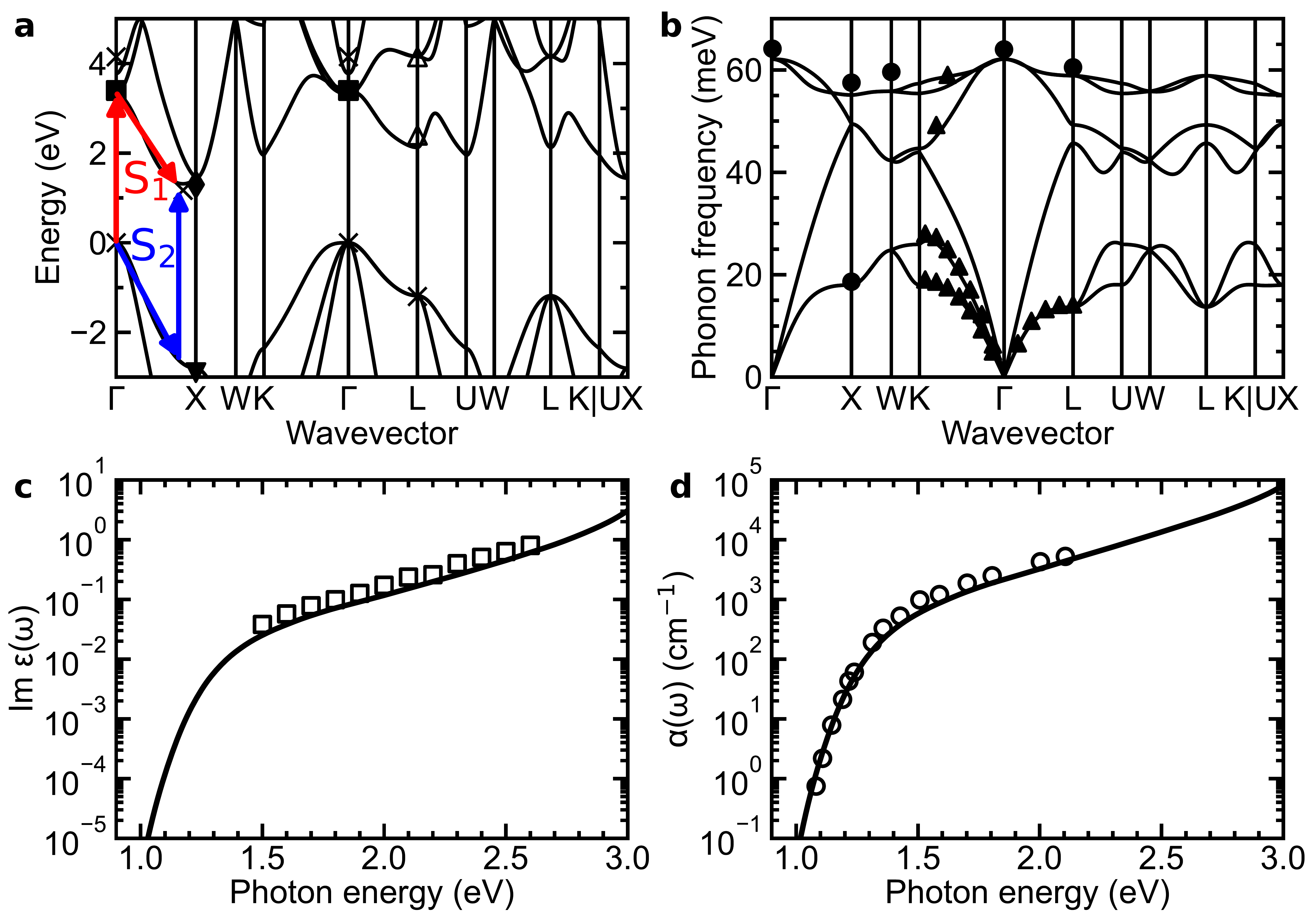}
    \caption{\textbf{Indirect phonon-assisted optical absorption spectra with \texttt{EPW}.}
    (a) calculated band structure of silicon including GW quasiparticle corrections (lines). The energy is referred to the top of the valence bands. The marks are measured energies of critical points: cross: Ref.~\cite{madelung1991o}; diamond: Ref.~\cite{Hybertsen1986s}; filled triangle: Ref.~\cite{spicer1968o}; filled square: Ref.~\cite{zucca1970o}; open triangle: Ref.~\cite{straub1985o}.
    (b) Calculated phonon dispersion relations of silicon (lines). Neutron scattering data are shown as disks ~\cite{dolling1963o} and filled triangles~\cite{nilsson1972o}.
    (c) Calculated imaginary part of the dielectric function of silicon (line) including phonon-assisted processes. The squares represent experimental data from Ref.~\cite{aspnes1983o}.
    (d) Calculated absorption coefficient of silicon. The circles are measurements from Ref.~\cite{tiedje1984o}.}
    \label{fig:optics}
\end{figure}

We evaluate quasiparticle corrections on a 6$\times$6$\times$6 $\mathbf{k}$-point grid within the GW method using the \texttt{BerkeleyGW} code~\cite{deslippe2012o}. The GW-corrected band gap is 1.31~eV, which slightly overestimates the experimental value of 1.12~eV at room temperature~\cite{sze2006o}, but agrees well with prior work using similar settings \cite{Hybertsen1986s}. The resulting quasiparticle band structure is shown in Fig.~\ref{fig:optics}(a). Our calculations agree well with experimental measurements at several critical points~\cite{Hybertsen1986s,wachs1985o,straub1985o,madelung1991o}, shown by marks in the figure. In Fig.~\ref{fig:optics}(a) we also show representative optical transitions leading to the amplitudes in Eq.~\eqref{eq:s1} and \eqref{eq:s2}. The fundamental gap is indirect, between the top of the valence band at the $\Gamma$ point and the bottom of the conduction band near the $X$ point. To correctly describe optical transitions near the fundamental gap, it is essential to include phonon-assisted processes.
Figure~\ref{fig:optics}(b) shows calculated phonon dispersion relations, which agree well with prior calculations by us~\cite{Ponce2016a} and other groups~\cite{petretto2018o}, as well as experimental data~\cite{dolling1963o,nilsson1972o} (solid symbols).
Figure~\ref{fig:optics}(c) shows the imaginary part of the dielectric function, as obtained from Eq.~\eqref{eqn:eps2} using 32$\times$32$\times$32 fine $\bk$- and $\bq$-point grids. The temperature is set to 300~K, and the Dirac delta functions in Eq.~\eqref{eqn:eps2} are approximated via Gaussians of width 50~meV. In this plot, a rigid shift of $-$0.19~eV is applied in order to match the GW band gap to the measured gap of silicon at room temperature (1.12~eV)~\cite{sze2006o}. 
Figure~\ref{fig:optics}(d) shows the related optical absorption coefficient from Eq.~\eqref{eq:abs}. This calculation requires the real part of the frequency-dependent refractive index, for which we used experimentally measured values from Ref.~\cite{schinke2015o} for simplicity. 

Our calculated spectra agree very well with experiments in the photon energy range between the indirect gap and the direct gap at 3.3 eV. Beyond the lineshape, the magnitude also agrees well with experiment over several orders of magnitude, with the theory underestimating the experimental data by 50\% at most. This residual underestimation might be related to the rigid shift of the band gap, which we did not include in the velocity renormalization expressed by Eq.~\eqref{eq:ren_vme}, and to the fact that electron-phonon matrix elements are slightly too weak in DFT as a result of the band gap problem~\cite{Giustino2017z,antonius2014o}.

  \subsection{Special displacement method}\label{sec.sdm}

\subsubsection{Background and formalism}

The \texttt{EPW} package contains standalone modules, the \texttt{ZG} toolset, for calculating finite-temperature properties including quantum zero-point effects via the special displacement method~\cite{Zacharias2016z, Zacharias2020z, Zacharias2021za}. The special displacement method is a supercell-based approach which is distinct from and complementary to the Wannier-Fourier interpolation method described in Sec.~\ref{sec.epwtheory} and employed in Secs.~\ref{sec.transport}-\ref{sec.indabs}. The founding principle of this method is that the effects of electron-phonon couplings on the electronic and optical properties of extended solids can be captured by performing calculations for a large supercell where the atoms have been displaced away from their equilibrium crystallographic sites. The displacements are chosen in such a way that the corresponding atomic configuration represents the best single-point approximant to the quantum thermal distribution of the atomic coordinates~\cite{Zacharias2020z}. 

Under the approximations of adiabatic Born-Oppenheimer decoupling and harmonic lattice, the quantum thermal average of an electronic or optical property described by the observable $O$ can be written as~\cite{Zacharias2020z}:
\begin{equation}\label{eq_SMD_1}
      O(T)  =
     \prod_{\bq\nu} \!\int\! \frac{dx_{\bq \nu}dy_{\bq \nu}}{\pi u^2_{\bq  \nu}} 
    e^{-|z_{\bq \nu}|^2/ u^2_{\bq  \nu}} O({\{\tau_{\kappa\alpha p}\}}), 
 \end{equation}
 where $\bq$ runs over the set of wavevectors in a uniform Brillouin zone grid which excludes time-reversal invariant points and time-reversal partners. In Eq.~\eqref{eq_SMD_1}, $z_{\bq \nu}$ denote normal mode coordinates with real part $x_{\bq \nu}$ and imaginary part $y_{\bq \nu}$. $u^2_{\bq \nu}  = (\hbar/2M_0 \omega_{\bq \nu}) (2n_{\bq  \nu} + 1 )$ is the mean-square displacement for the oscillator $\bq\nu$ with Bose-Einstein occupation $n_{\bq  \nu}$; $O({\{\tau\}})$ is the property of interest, such as for example the Kohn-Sham eigenvalues, density of states, or optical absorption spectrum, calculated for the set of atomic coordinates $\{\tau_{\kappa\alpha p}\}$. The relation between these coordinates and the normal coordinates $z_{\bq\nu}$, which is required to carry out the integral in Eq.~\eqref{eq_SMD_1}, is provided in Ref.~\cite{Zacharias2020z}. 

In the special displacement method, the configurational average expressed by Eq.~\eqref{eq_SMD_1} is approximated by a single calculation for an optimum configuration:
\begin{equation}\label{eq_SMD_2}
     O(T) \simeq O({\{\tau^0_{\kappa\alpha p} + \Delta \tau^{\rm ZG}_{\kappa\alpha p}}\}),  
\end{equation}
where $\tau^0_{\kappa\alpha p}$ represent atomic coordinates in the DFT ground-state at zero temperature, and the optimum ``ZG'' displacement is given by~\cite{Zacharias2020z}: 
 \begin{equation}\label{eq.realdtau_method00}
   \DD\tau^{\rm ZG}_{ \k\alpha p} =  \sqrt{\frac{M_0}{N_{p} M_\k}} 2\!\sum_{\bq\nu}\! 
   S_{\bq  \nu} u_{\bq \nu}  \, {\rm Re}
   \Big[ e^{i\bq \cdot {\bf R}_p} e_{\k\alpha,\nu} (\bq ) \Big] . 
 \end{equation}
As in Eq.~\eqref{eq_SMD_1}, the summation is restricted to $\bq$-points from a uniform Brillouin zone grid which are not time-reversal invariant and are not time-reversal partners. This partitioning is described in Appendix~B of Ref.~\cite{Giustino2017z}. The quantities $S_{\bq  \nu}$ appearing in Eq.~\eqref{eq.realdtau_method00} are signs ($\pm 1$) determined by the \texttt{ZG} module so as to guarantee that the resulting displacements make $O({\{\tau^0_{\kappa\alpha p} + \Delta \tau^{\rm ZG}_{\kappa\alpha p}}\})$ the best possible approximant to Eq.~\eqref{eq_SMD_1}.
In the thermodynamic limit of a large supercell, the ZG displacements reproduce the exact mean-square anisotropic displacement tensors, which are given by~\cite{Bruesch1982z}:
 \begin{equation}\label{eq.mean-square_disp}
   U_{\k,\a\a'}(T) = \frac{\hbar}{2 M_\k\omega_{\bq\nu}} \sum_\nu\int\!\frac{d\bq}{\Omega_{\rm BZ}}\, 
   e_{\k \a, \nu } (\bq) e^{*}_{\k \a',\nu } (\bq) \, (2n_{\bq \nu}+1) .
 \end{equation}
In the same limit, a single evaluation of the property $O({\{\tau^0_{\kappa\alpha p} + \Delta \tau^{\rm ZG}_{\kappa\alpha p}}\})$ tends to the exact thermal average in Eq.~\eqref{eq_SMD_1}. In the case of non-periodic systems, such as for example nanocrystals and quantum dots, these equations are replaced by their $\Gamma$-point only versions~\cite{Zacharias2020zb,Zacharias2021zb,Zacharias2016z}.

The special displacement method is similar in spirit to computing thermodynamic averages using path-integral molecular dynamics~\cite{Ramirez2006z}, but it differs insofar a single calculation is required to evaluate the average instead of many molecular dynamics snapshots. In this method, electron-phonon couplings are included non-perturbatively through the changes of the Kohn-Sham energies and wavefunctions caused by the ZG displacements. The method can be applied to compute any property that can be expressed by means of a Fermi Golden Rule, such as for example temperature-dependent band structures, density of states, and optical spectra. The main limitation of this approach as compared to the strategy outlined in Secs.~\ref{sec.transport}-\ref{sec.indabs} is that, being an adiabatic theory, fine spectral features on the scale of the phonon energy are averaged out. Conversely, its main advantage is that it is easy to use as it requires a single DFT calculation. 

The Kohn-Sham energy eigenvalues generated by the special displacement method capture temperature renormalization and quantum zero-point effects at the same level as the adiabatic Allen-Heine theory of temperature-dependent band structures~\cite{Allen1976z}. The calculation of the imaginary part of the dielectric function including temperature effects, zero-point corrections, and phonon-assisted indirect processes, is performed by evaluating:
\begin{equation}\label{eq.eps2}
  {\rm Im}[\varepsilon(\omega)] = \frac{2 \pi e^2}{ \varepsilon_0 m_{\rm e}^2 \Omega_{\rm sc}} 
      \frac{1}{\,\omega^2}
  \sum_{cv} \int\frac{d\bf K}{\Omega_{\rm BZ,sc}}
     | \braket{\psi_{c {\bf K}} | {\mathbf e}\cdot \hat{{\bf p}} | \psi_{v{\bf K}}}|^2 \delta(\varepsilon_{c {\bf K}}-\varepsilon_{v {\bf K}}-\hbar\omega),
\end{equation}
where $\Omega_{\rm sc}$ and $\Omega_{\rm BZ,sc}$ are the volumes of the supercell and the supercell Brillouin zone, respectively, the summations over $v,c$ refer to valence and conduction states, and $\bf K$ is a wavevector of the supercell Brillouin zone. This expression contains both direct and indirect optical transitions: the indirect transitions are hidden in the dependence of the wavefunctions $\psi_{v{\bf K}}$ and $\psi_{c{\bf K}}$ on the ZG displacement. Similarly, this expression includes temperature renormalization via the dependence of the energies $\varepsilon_{v {\bf K}}$  and $\varepsilon_{c {\bf K}}$ on the ZG displacements~\cite{Zacharias2016z}.

\subsubsection{Computational considerations}

Calculations using the special displacement method proceed as follows. First, one performs phonon calculations for the crystalline unit cell (not the supercell) using standard DFPT on a coarse uniform Brillouin zone grid. From this calculation, the phonon frequencies and eigenmodes are computed on a finer Brillouin-zone grid with $N_1\!\times\! N_2\!\times \!N_3$~$\bq$-points using standard interpolation of the force constant matrix~\cite{Baroni2001s}. Then the ZG displacement within a supercell consisting of $N_1\!\times\! N_2\!\times \!N_3$ unit cells is evaluated at the temperature $T$ via Eq.~\eqref{eq.realdtau_method00}. The desired property is finally computed with this supercell, with the atoms displaced according to $\DD\tau^{\rm ZG}_{ \k\alpha p}$. The special displacements are generated by the \texttt{ZG} module; the procedure is computationally inexpensive and is performed serially.

The \texttt{ZG} toolset provides several codes to analyze the results of supercell calculations performed with special displacements, for example Brillouin-zone unfolding of temperature-dependent band structures from the supercell to the unit cell, density of states, electronic spectra, and vibrational spectra. 

Band unfolding is performed using the procedure outlined in Ref.~\cite{Popescu2012z}, which consists of determining the spectral function in the primitive Brillouin zone by projecting the wavefunctions of the supercell into the wavefunctions of the unit cell. In practice, the spectral density for the wavevector $\bk$ at the energy $\varepsilon$ is obtained as: 
\begin{equation} \label{eq.spctrl_fn}
  A_{\bf k}(\varepsilon) = \sum_{m {\bf K}} P_{m {\bf K},{\bf k}} \,\delta ( \varepsilon - \varepsilon_{m {\bf K}}),
\end{equation}
where the spectral weights $P_{m {\bf K},{\bf k}}$ are given by:
\begin{equation}\label{eq.sprtl_weight_T2}
  P_{m\bK,\bk} = \sum_{\bG_{\rm sc}}|c_{m \bK} ({\bf G}_{\rm sc} + \bk - \bK) |^2.
\end{equation}
Here, $c_{m \bK}$ denote planewaves coefficients of supercell wavefunctions, and $\bG_{\rm sc}$ denote reciprocal lattice vectors of the supercell. Equation~\eqref{eq.sprtl_weight_T2} refers to norm-conserving pseudopotential implementations. The \texttt{ZG} module includes additional terms that are required in this expression when using ultrasoft and PAW (projector augmented-wave) pseudopotentials~\cite{Zacharias2020z,Zacharias2021zb}. A similar unfolding strategy is employed to analyze lattice dynamics, as well as X-ray and neutron diffuse scattering intensities accounting for multiphonon interactions~\cite{Zacharias2021zc}.

\subsubsection{Application example}

To demonstrate the implementation of the \texttt{ZG} module, we investigate the temperature-dependent band structure renormalization and phonon-assisted optical absorption spectra of silicon and BaSnO$_3$. We perform calculations using the local density approximation (LDA) for the exchange and correlation~\cite{Perdew1981z,Ceperley1980z} and ONCV pseudopotentials~\cite{Hamann2013s,Schlipf2015s}. We use a planewaves kinetic energy cutoff of 40~Ry for silicon and 120~Ry for BaSnO$_3$, and 6$\times$6$\times$6 uniform {\bf k}-point grids for ground-state calculations. With these settings, we obtain indirect and direct band gaps of 0.49~eV and 2.56~eV for silicon, respectively, and indirect and direct band gaps of 1.06~eV and 1.55~eV for BaSnO$_3$, respectively. We perform calculations of dynamical matrices on 4$\times$4$\times$4 {\bf q}-point grids in both cases, and use the \texttt{ZG} module to generate displacements in 3$\times$3$\times$3 supercells. The signs $S_{\bq\nu}$ appearing in Eq.~\eqref{eq.realdtau_method00} are determined by minimizing the error descriptor in Eq.~(54) of Ref.~\cite{Zacharias2020z} with a dimensionless threshold $\eta = 0.1$, after enforcing a smooth Berry connection between vibrational eigenmodes across the Brillouin zone~\cite{Zacharias2020z}. Kohn-Sham energies in the structures with ZG displacements are calculated using a 12$\times$12$\times$12 uniform $\bK$-point grid in the supercell Brillouin zone, and 108 unoccupied states. For calculations of dielectric functions, we use up to 200~randomly-generated ${\bf K}$-points in the supercell Brillouin zone, as well as 27 and 135 conduction bands for silicon and BaSnO$_3$, respectively. All Dirac deltas are replaced by Gaussians of width 30~meV.

\begin{figure*}[htb!]
\includegraphics[width=0.99\textwidth]{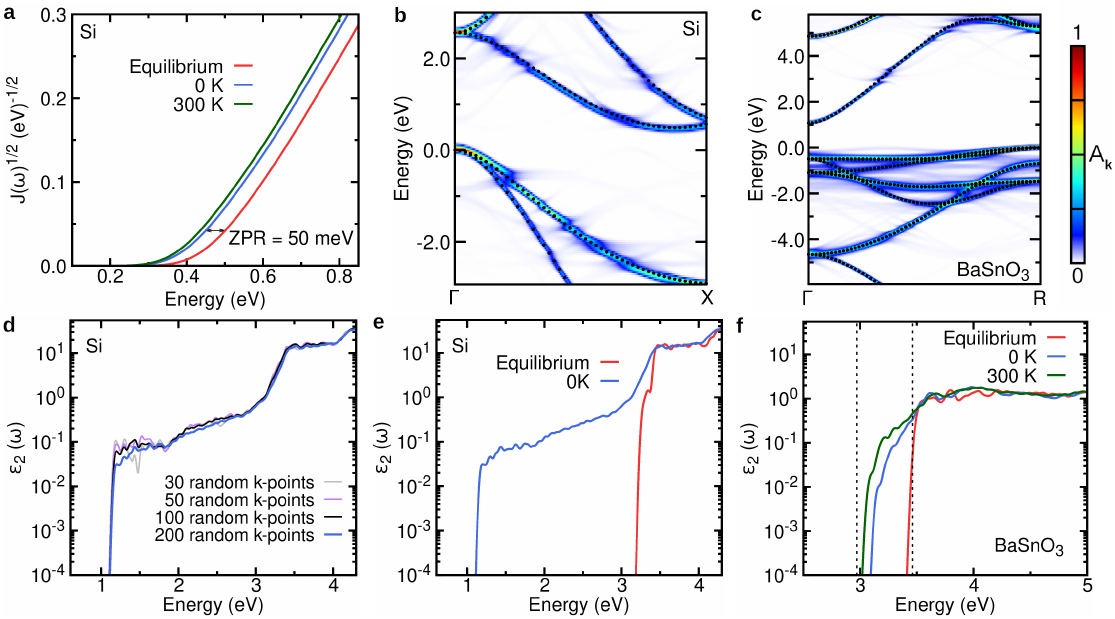}
 \caption{\textbf{Applications of the special displacement method implemented in the \texttt{ZG} module.}
  (a) Square root of the JDOS of silicon evaluated with atoms in the DFT ground-state structure (red); using special displacements at 0\,K (blue); and using special displacements at 300~K (green). The red-shift of the JDOS with respect to the ground-state structure signals the quantum zero-point correction to the band gap (0\,K curve) and its temperature renormalization (300\,K curve), respectively. (b) Electron spectral function of silicon calculated with the special displacement method at 0\,K, along the $\Gamma$X path in the Brillouin zone. The black disks indicate the band structures calculated in the crystalline unit cell with the DFT ground-state geometry. (c) Electron spectral function of BaSnO$_3$ calculated with the special displacement method at 0\,K, along the $\Gamma$R path in the Brillouin zone. The black disks indicate the band structures calculated in the crystalline unit cell with the DFT ground-state geometry. 
  (d) Convergence of the imaginary part of the dielectric function of silicon at 0\,K with respect to the number of random ${\bf K}$-points used to sample the Brillouin zone of the supercell. (e) Imaginary part of the dielectric function of silicon calculated with the atoms in the DFT ground-state structure (red), and by using special displacements at 0\,K (blue). A scissor shift of 0.64~eV is employed to match the experimental gap~\cite{Green1995z}.
  (f) Imaginary part of the dielectric function of BaSnO$_3$ calculated with the atoms in the DFT ground-state structure (red), and by using special displacements at 0\,K (blue) and 300\,K (green). The vertical dashed lines indicate the direct and indirect band gap energies. 
  A scissor shift of 1.91~eV is employed to match the experimental gap~\cite{Kim2012z}.
 \label{fig_ZG} }
\end{figure*}

Figure~\ref{fig_ZG}(a) shows the joint density of states (JDOS) of silicon at 0~K (red) and 300~K (blue), as calculated using the special displacement method. For comparison the JDOS computed for the DFT ground state structure is also shown in green. We see that, upon including electron-phonon interactions via the special displacements, even at 0~K the JDOS is red-shifted with respect to the DFT ground state. This is a manifestation of the zero-point band gap renormalization~\cite{Giustino2010z}. Upon increasing temperature, the onset of the JDOS further red-shifts. This temperature-induced band gap narrowing is referred to as Varshni effect~\cite{Varshni1967z}. The horizontal offset between the JDOS for the DFT ground state structure and the JDOS computed with special displacements at 0~K yields a zero-point renormalization of 50~meV, in good agreement 
with prior work~\cite{Karsai2018z,Ponce2015z}. More accurate values can be calculated by increasing the supercell size~\cite{Zacharias2016z,Zacharias2020z}. The band gap renormalization can also be computed without using the JDOS; to this end, one needs to evaluate the Kohn-Sham eigenvalues at the supercell $\bK$-points that unfold onto the wavevectors of the band extrema in the Brillouin zone of the unit cell. 

Figures~\ref{fig_ZG}(b) and (c) report color maps of the electronic spectral functions of silicon and BaSnO$_3$ at 0~K calculated using the special displacement method. For comparison, the band structures in the DFT ground state are overlaid to these color maps. These maps can directly be compared to angle-resolved photoelectron spectroscopy data. From these images, we identify numerically the quasiparticle band structures by extracting the spectral peaks. In the case of silicon, the valence band maximum blue-shifts by 32~meV with respect to ground-state DFT, and the conduction band red-shifts by 18~meV. The resulting gap renormalization of 50~meV is in agreement with the value determined in Fig.~\ref{fig_ZG}(a) via the JDOS. In the case of BaSnO$_3$, Fig.~\ref{fig_ZG} shows a zero-point renormalization of the band gap of 10~meV; however, we emphasize that this value is not fully converged, and larger supercells as well as corrections for Fr\"ohlich couplings~\cite{Nery2016z} are necessary to obtain accurate data.

Figure~\ref{fig_ZG}(d) reports a convergence test of the imaginary part of the dielectric function of silicon, as computed with the special displacement method. In this calculation, we keep the supercell fixed, and we increase the number of random {\bf K}-points in the supercell Brillouin zone. It is seen that the dielectric function converges relatively rapidly with the number of points, and full convergence is achieved with 200 points when the Dirac delta functions in Eq.~\eqref{eq.eps2} are replaced by Gaussians with 30~meV width. Clearly, a larger smearing would require fewer {\bf K}-points. 

In Fig.~\ref{fig_ZG}(e) we compare the imaginary part of the dielectric function of silicon computed in the DFT ground state structure (red) with the special displacement method (blue). We see that the special displacements correctly capture phonon-assisted indirect optical transitions in the energy range between the indirect gap and direct gap of silicon. This approach provides an alternative strategy for computing optical spectra to the indirect optics module of \texttt{EPW} described in Sec.~\ref{sec.indabs}. In the special displacement method, both temperature-dependent band structure renormalization and phonon-assisted processes are included on the same footing.

Figure~\ref{fig_ZG}(f) shows the imaginary part of the dielectric function of BaSnO$_3$ at 0\,K (blue) and 300\,K (green), including phonon-assisted processes. For comparison, we also show the calculation using the DFT ground-state structure (red), which misses phonon-assisted transitions. The spectra compare well with prior work using the special displacement method~\cite{Kang2018z}.

\section{Implementation and HPC benchmarks}\label{sec.hpc}

In this section we describe some of implementation and parallel programming models of \texttt{EPW v6}. We first describe the computational workflow and basic capabilities of \texttt{EPW v6} (Sec.~\ref{sec.hpc1}). Then we outline our recent efforts to prepare \texttt{EPW} for exascale HPC systems with a focus on a newly implemented highly-scalable parallelization scheme, hybrid two-level MPI and OpenMP parallelization. We present benchmarking results which demonstrate that, with this new parallelization strategy, \texttt{EPW v6} can run at nearly full scale on the pre-exascale HPC system Frontera at the Texas Advanced Computing Center (TACC) (Sec.~\ref{sec.hpc2}). Finally, we describe the I/O strategy employed in \texttt{EPW v6}, which is based on XML (eXtensible Markup Language) and HDF5 file formats (Sec.~\ref{sec.hpc3}). We emphasize that \texttt{EPW} continues to evolve with changes and improvements, and the reader is referred to the \texttt{EPW} website for its most up-to-date features and functionalities~\cite{EPW-website}.

\subsection{Computational workflow and basic capabilities}\label{sec.hpc1}

Starting from version 6, the \texttt{EPW} code is divided into two separate executables, \texttt{pw2epw.x} and \texttt{epw.x}. \texttt{pw2epw.x} serves as an interface to the \texttt{PWscf} and \texttt{PHonon} codes of \texttt{Quantum ESPRESSO}. The code imports ground-state charge density with computational parameters from the prior self-consistent run with \texttt{pw.x}, and dynamical matrices, variations of the Kohn-Sham potentials, and vibrational mode pattern files from a prior phonon calculation with \texttt{ph.x}. Then \texttt{pw2epw.x} performs non-self-consistent
calculations to obtain Kohn-Sham wavefunctions on a coarse {\bk}-point grid, and it calls \texttt{Wannier90} in library mode to calculate the unitary rotation matrices $U_{mn\bk}$ in Eq.~\eqref{eq.wannier-def} needed to transform Bloch states into maximally localized Wannier functions. Subsequently, \texttt{pw2epw.x} evaluates electron-phonon matrix elements on coarse $\bk$- and $\bq$-point grids, and it transforms the Hamiltonian, dynamical matrices, and electron-phonon matrix elements into the Wannier representation. Finally, it ends with outputting relevant quantities in XML and HDF5 formats for subsequent runs with \texttt{epw.x}. The main program of \texttt{EPW}, \texttt{epw.x}, carries out Wannier interpolation of the Hamiltonian, dynamical matrices, and electron-phonon matrix elements on fine $\bk$- and $\bq$-point grids after reading these quantities in the Wannier representation, and then it calculates electron-phonon related properties on these grids.

The current snapshot of the code supports spin-unpolarized and
non-magnetic spin-orbit calculations; work is currently in progress to extend the code to the cases of spin-polarized and non-collinear magnetic cases. Regarding pseudopotentials (PPs), norm-conserving (NC) PPs are supported. Since both ultrasoft PPs and the PAW method violate fundamental symmetry relations that the electron-phonon matrix elements must fulfill~\cite{Engel2022}, extending the current NCPP implementation to these cases poses some challenges. We will proceed to these extensions in future releases.

\subsection{Parallelization}\label{sec.hpc2}

At the time of writing of this manuscript, we witness the launch of the first supercomputer entering the exascale computing era, namely Frontier at Oak Ridge National Laboratory; additional exascale systems are currently being deployed worldwide. Exascale HPC systems are characterized by many-core and heterogeneous architectures. Since 2020, we have made efforts to ready \texttt{EPW} for the exascale transition. In the following, we report on one of the outcomes of this ongoing effort, a hybrid two-level MPI/OpenMP parallelization scheme, which makes \texttt{EPW} work seamlessly and efficiently on many-core architectures.

Previous versions of \texttt{EPW} adopted one-level parallelization 
over the electron wavevectors ($\bk$-point parallelization) via MPI. This strategy has three shortcomings, which lead to an early saturation of the speedup with the number of cores in large-scale calculations: (i) There is an upper bound for the number of total MPI tasks which guarantees good scalability; this bound is determined by the number of $\bk$-points. (ii) As the number of MPI tasks increases, the overhead cost associated with MPI communications, in particular collective communications, increases. (iii) Due to the internal buffers of MPI, the memory overhead also increases with the number of MPI tasks. To address these issues, we extended the one-level parallelization over $\bk$-points to the hybrid two-level MPI and OpenMP parallelization over both $\bk$- and $\bq$-points.

\begin{figure}
    \includegraphics[width=\textwidth]{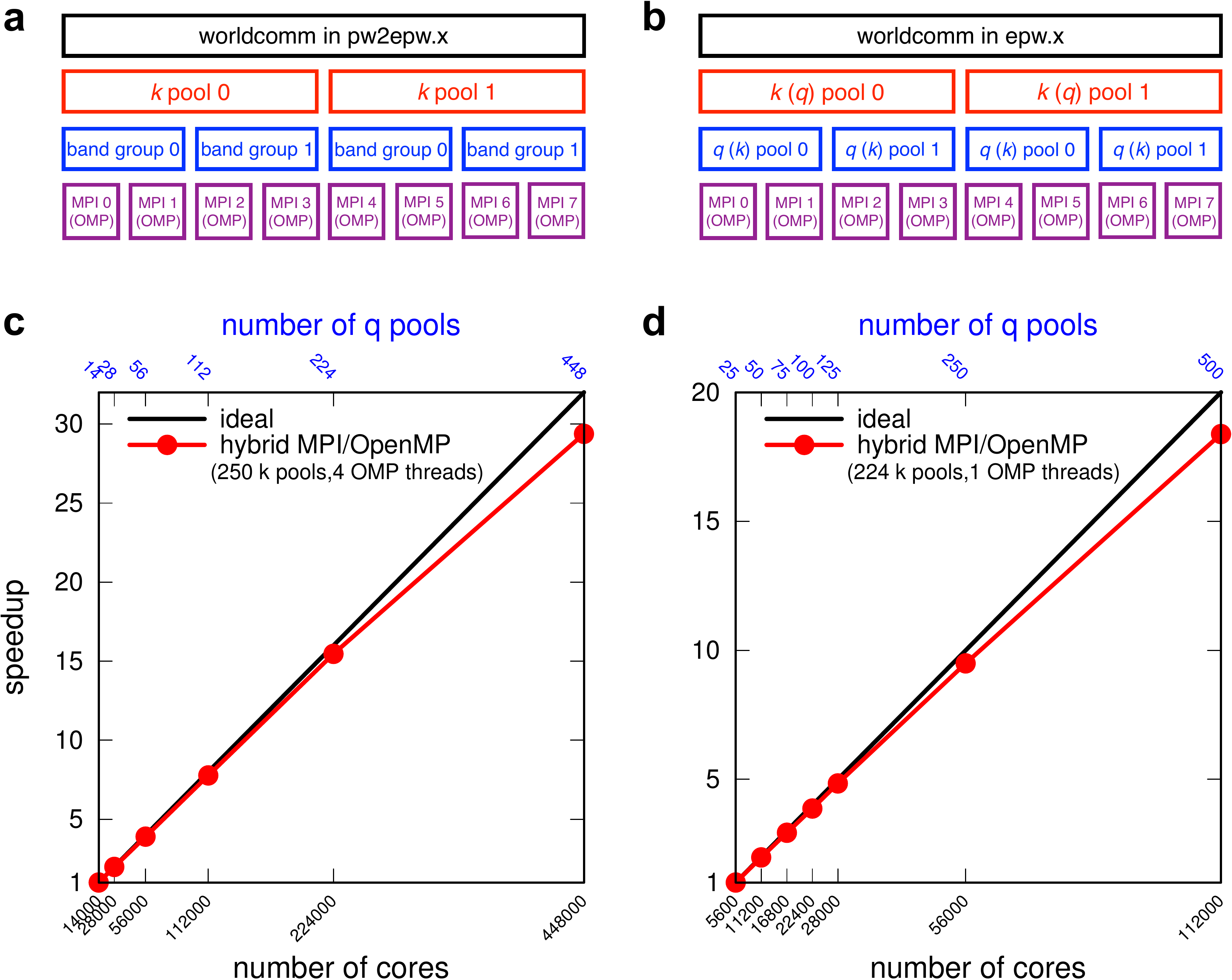}
    \caption{\textbf{Parallelization strategy and scaling benchmarks of \texttt{EPW v6}.} (a) Schematic of hybrid two-level MPI and OpenMP parallelization employed in \texttt{pw2epw.x}. In this example, we consider a hypothetical scenario with 8 MPI tasks in total. Processes are divided into two upper MPI groups. Each group is further divided into two lower MPI groups, each consisting of a set of MPI tasks. OpenMP is employed within each MPI task. (b) Schematic of hybrid two-level MPI and OpenMP parallelization employed in \texttt{epw.x}. (c) Strong-scaling test for the evaluation of the electron-phonon matrix elements of MgB$_2$. (d) Strong-scaling tests for the solution of the anisotropic Eliashberg equations for MgB$_2$.
    \label{fig_par}}
\end{figure}

In the hierarchical two-level MPI and OpenMP parallelization strategy, the total MPI tasks are partitioned in two levels of MPI groups, each consisting of a set of MPI tasks; in the lowest level, OpenMP parallelization is employed within each MPI task. For instance, in the \texttt{pw2epw.x} code, the total MPI tasks are first divided into $\bk$-point pools, and each pool is further divided into band groups. Each band group contains a set of MPI tasks, and OpenMP parallelization is employed within each MPI task, as shown in Fig.~\ref{fig_par}(a).
In the \texttt{epw.x} code, the total MPI tasks are first divided into $\bk$- or $\bq$-point pools, and each pool is further divided into $\bq$- and $\bk$-point pools, respectively. Each $\bq$- or $\bk$-point pool contains a set of MPI tasks, and  OpenMP parallelization is employed within each MPI task. This is shown in Fig.~\ref{fig_par}(b). The parallelization option employed in the upper-level pool (over $\bk$-points or $\bq$-points) is chosen so as to maximize the calculation efficiency. Currently, $\bq$-point parallelization in the upper-level pool is used in phonon self-energy calculations, while $\bk$-point parallelization in the upper-level pool is used in all other scenarios.

As compared to the one-level parallelization, the hierarchical two-level MPI and OpenMP parallelization can reduce the number of MPI tasks in each MPI communicator group, thereby reducing the overhead cost for collective MPI communication as well as the memory footprint due to the internal buffers of the MPI library. OpenMP parallelization can further reduce the memory footprint associated with replicated memory allocations in the code. This strategy enables increased flexibility in distributing the computational workload over a large number of cores.

To demonstrate the scaling behavior of \texttt{EPW v6}, we perform strong-scaling tests for the two-gap superconductor MgB$_2$. We evaluate electron-phonon matrix elements on 150$\times$150$\times$150 $\bk$- and $\bq$-point grids [Fig.~\ref{fig_par}(c)], and we solve the anisotropic Eliashberg equations on 72$\times$72$\times$72 $\bk$ and $\bq$ grids [Fig.~\ref{fig_par}(d)]. All calculations are performed on the Frontera supercomputer at TACC. Frontera consists of 8,368 nodes, each equipped with two Intel Xeon Platinum 8280 (``Cascade Lake'') processors with 28 cores per processor. To minimize statistical fluctuations in these benchmarks, we execute each calculation four times, and we average the resulting Wall times. Figures~\ref{fig_par}(c) and (d) show that, using this new parallelization scheme, we achieve approximately 92\% of the ideal speedup up to 448,000 cores in the evaluation of the electron-phonon matrix elements, and approximately 92\% of the ideal speedup up to 112,000 cores in the solution of the anisotropic Eliashberg equations.

The hybrid two-level MPI and OpenMP parallelization scheme is controlled by three parameters that can be optimized to achieve maximum parallel scaling efficiency: the number of upper-level MPI tasks, the number of lover-level MPI tasks, and the number of OpenMP threads per MPI task. In the benchmarks shown in Figs.~\ref{fig_par}(c) and (d), we fix the number of {\bk}-point pools to 250 and 224, respectively, by varying
the number of $\bq$-point pools depending on the number of cores. With these
choices, we are able to achieve near-ideal speedup on extreme scaling tests; we expect that a speedup even superior to 92\% could be achieved upon further optimization of the parallelization parameters. In the case of the solution of the Eliashberg equations, we fix the number of iterations to 40 in order to have a meaningful comparison between runs using different settings and number of cores. We also point out that, in large-scale runs, I/O time dominates the total Wall time. To overcome this I/O bottleneck, we employ a low-I/O mode whereby memory is exploited as much as possible, and minimum usage of storage media is made.

Our strategy for OpenMP parallelization is to use threaded versions of scientific libraries such as Intel MKL and Cray LibSci, rather than explicitly using OpenMP directives throughout the code. For the benchmarks reported in Fig.~\ref{fig_par}(c), we find that exceeding 7 OpenMP threads per MPI task worsens the performance. 
Therefore we use 4~OpenMP threads per MPI task, and we make all threads belonging to the same MPI task reside in the same NUMA (non-uniform memory access) domain. Additionally, we carried out code optimization at the individual node level. For instance, we changed the order of DO loops in favor of improved data locality and cache reuse in computationally-intensive parts such as the superconducting module. 

Another significant improvement in \texttt{EPW v6} is the more extensive use of crystal symmetry operations on the coarse $\bk$- and $\bq$-point grids. In previous versions of the code, symmetry was used to generate the variations of the Kohn-Sham potentials and the dynamical matrices, which are read from the \texttt{PHonon} code, from the irreducible wedge of the Brillouin zone to the full zone~\cite{Giustino2007l}. In \texttt{EPW v6}, users have the option to directly  rotate Kohn-Sham wavefunctions on the coarse $\bk$-point grid, and to use these wavefunctions in the evaluation of electron-phonon matrix elements for the star of each irreducible $\bq$-point, the overlap and projection matrices for Wannierization, and the velocity matrix elements. For example, the generation of electron-phonon matrix elements from the irreducible wedge of the Brillouin zone to the full zone is achieved as follows (using the convention of active transformations~\cite{Bradley2009}):
\begin{equation} \label{eq:sym.eph}
  g_{mn\nu}({\bf k},S{\bf q})=\sum_{m'n'}\langle\psi_{m'S^{-1}{\bf k}+{\bf q}}({\bf r})|\psi_{m{\bf k}+S{\bf q}}(\{S|{\bf v}\}{\bf r})\rangle^*g_{m'n'\nu}(S^{-1}{\bf k},{\bf q})\langle\psi_{n'S^{-1}{\bf k}}({\bf r})|\psi_{n{\bf k}}(\{S|{\bf v}\}{\bf r})\rangle\,,
\end{equation}
where $S$ is the rotation part of crystal symmetry operation, $\bf v$ is the fractional translation for non-symmorphic space groups, and the summation is over each degenerate subspace.

In many applications it is useful to also exploit symmetry operations on the fine grids, so as to reduce computational cost. Here a word of caution is needed: since the Wannierization procedure does not impose any symmetry, quantities on the fine grids are not expected to fulfill the required symmetry relations. A correct use of symmetry reduction on the fine grids would require the direct generation of symmetric Wannier functions~\cite{Sakuma2013} or to recover the symmetry of Wannier functions after Wannierization and before interpolation~\cite{Gresch2018}.

\subsection{Structured I/O}\label{sec.hpc3}

\texttt{EPW v6} employs the XML and the HDF5 data formats. Both formats can describe hierarchical data, are self-describing, flexible, and portable. For small to medium-sized data, the XML format is used; for large binary data, the HDF5 file format is used in parallel. In previous versions of \texttt{EPW}, several large arrays were read or written via per-process I/O or parallel MPI I/O. The former quickly overwhelms the file systems in large-scale calculations by generating a large number of files; the latter is not easily portable. We addressed both of these issues for large-size binary data by employing parallel I/O via parallel HDF5. 

We note that parallel I/O performance heavily depends on the details of file systems and the parallelization settings. For instance, in the case of the LUSTRE file system, the number and size of stripes play an essential role in achieving high performance. In the case of the ROMIO implementation of MPI-IO, identifying optimum parameters in large-scale runs is nontrivial and requires systematic experimentation on HPC systems.

\section{Future directions}\label{sec.futuredirections}

Since the last technical paper in 2016 \cite{Ponce2016a}, the \texttt{EPW} code has considerably expanded in scope, to the point of becoming a robust and efficient software platform for developing, testing, and deploying new methods addressing electron-phonon physics and related materials properties. It is then natural to ask which components and functionalities are still in need of improvement, what are emerging new directions in this area, and more generally what comes next for the \texttt{EPW} project.

On the functionality front, we expect to continue expanding the capabilities for transport, superconductivity, polarons, and optics described in Sec.~\ref{sec.functionalities}. For example, the transport module is based on the steady-state linearized Boltzmann transport equation; here, it would be desirable to generalize the methodology to high-field transport and to time-dependent driving fields. Similarly, in the superconductivity module, the treatment of electron-electron repulsion could be improved by seamlessly integrating \texttt{EPW} with standard GW codes. The study of polaron physics from an \textit{ab initio} many-body perspective is just beginning, therefore we can expect a number of developments in this area, from the study of polaron transport to their optical properties and their relation with the theory of band structure renormalization. Calculations of phonon-assisted optical transitions are currently restricted to optical absorption spectra, but the formalism can just as well be employed for investigating related phenomena such as for example Auger-Meitner recombination~\cite{Bushick2022}. Beyond these core modules of the \texttt{EPW} project, we anticipate growth in the calculation of electronic and optical properties using the special displacement method within the \texttt{EPW/ZG} code; for example, the current implementation focuses on harmonic systems, but generalizations to strongly anharmonic systems have recently been proposed~\cite{Zacharias2022f}.

On the accuracy front, we anticipate that future work will focus on improving the precision of the Wannier interpolation method that is the basis for \texttt{EPW}. For example, new methods that take dimensionality into account in the interpolation of the electron-phonon matrix elements will be essential to perform predictive and reliable calculations of transport, optics, superconductivity, and polarons in 2D materials~\cite{Sio2022l,Ponce2022lb}. Furthermore, it would be highly desirable to improve the predictive power of the DFPT electron-phonon matrix elements before even proceeding to Wannier-Fourier interpolation; a recent proposal to employ the GW method to calculate many-body corrections to the DFPT matrix elements offers a promising path toward this goal~\cite{ZhengluLi2022f}.

On the front of HPC, we envision continuing the current effort to enable \texttt{EPW} for exascale computing architectures. In addition to the hybrid MPI/OpenMP parallelization described in Sec.~\ref{sec.hpc}, it will be necessary to leverage new and diverse GPU architectures, and to enable the code for large-scale runs that will become possible with the new generation of exascale supercomputers that are being deployed worldwide. 

Another important direction will be to enhance the interoperability of \texttt{EPW} with other major electronic structure software packages such as \texttt{Abinit}~\cite{Abinit2020}, \texttt{VASP}~\cite{VASP1993}, \texttt{Siesta}~\cite{Siesta2002}, and \texttt{GPAW}~\cite{Enkovaara2010}. As several codes currently use the information generated by \texttt{EPW} for a variety of post-processing tasks, it will make sense to develop standardized data structures that adhere to the FAIR data principles. 

Regarding the programming model, transitioning toward object-oriented programming (OOP) will be useful due to the potential for high modularity, extensibility, and reusability, which will facilitate flexible and sustained software development~\cite{Gamma1995}. We expect that a targeted use of OOP will greatly increase the efficiency and productivity in the development and maintenance of the code~\cite{ismailbeigi2000}.

Alongside these developments, we anticipate increased attention to the issue of automation and the capability of performing electron-phonon calculations at scale for high-throughput approaches, data science, and artificial intelligence/machine learning (AI/ML) applications.

\section{Conclusions}\label{sec.conclusions}

In this manuscript we provided a comprehensive update of the current status, functionalities, and performance of the \texttt{EPW} code, a software project for \textit{ab initio} calculations of electron-phonon interactions and related materials properties. We described new algorithmic developments and calculation capabilities that have been introduced in the code since the previous status update in 2016~\cite{Ponce2016a}. In particular, in Sec.~\ref{sec.methodology} we outlined the methodological basis of the code and recent developments for computing electron-phonon matrix elements with high accuracy. In Sec.~\ref{sec.functionalities} we described new computational capabilities that are available with the current release \texttt{EPW v6}: calculations of carrier transport under both electric and magnetic fields within the \textit{ab initio} Boltzmann transport equation, including both carrier-phonon and carrier-impurity scattering; calculations of the superconducting gap function and critical temperature using the full-bandwidth anisotropic Eliashberg equations; calculations of wavefunctions and formation energies of both small and large polarons using the \textit{ab initio} polaron equations; calculations of optical absorption spectra including phonon-assisted indirect transitions; and calculations of temperature-dependent electronic and optical properties using the special displacement method. In Sec.~\ref{sec.hpc} we described our efforts to refactor the code in preparation for the exascale transition, and in particular to enable \texttt{EPW} for large-scale runs on massively-parallel supercomputers. Finally, in Sec.~\ref{sec.futuredirections} we offered an overview of possible future directions for this and similar codes. We hope that, beyond capturing a snapshot of the current status of the \texttt{EPW} project, the present manuscript will contribute to making advanced electron-phonon calculations more accessible and more widely used in computational materials discovery and design.

\begin{acknowledgments}
This research is supported by: the Computational Materials Sciences Program funded by the U.S. Department of Energy, Office of Science, Basic Energy Sciences, under Award No. DE-SC0020129 (project coordination, scale-up, polaron module, transport module, optics module, special displacement module); the National Science Foundation, Office of Advanced Cyberinfrastructure and Division of Materials Research under Grants No. 2103991 and 2035518 (superconductivity module, interoperability); the NSF Characteristic Science Applications for the Leadership Class Computing Facility program under Grant No.~2139536
(preparation for LCCF); the Fond National de la Recherche Scientifique of Belgium (F.R.S.-FNRS) and the European Union’s Horizon 2020 research and innovation program under grant agreements No. 881603-Graphene Core3 (transport module).
This research used resources of the National Energy Research Scientific Computing Center and the Argonne Leadership Computing Facility, which are DOE Office of Science User Facilities supported by the Office of Science of the U.S. Department of Energy, under Contracts No. DE-AC02-05CH11231 and DE-AC02-06CH11357, respectively.
The authors acknowledge the Texas Advanced Computing Center (TACC) at The University of Texas at Austin for providing access to Frontera, Lonestar6, and Texascale Days, that have contributed to the research results reported within this paper (http://www.tacc.utexas.edu); the Extreme Science and Engineering Discovery Environment (XSEDE)~\cite{XSEDE} which is supported by National Science Foundation grant number ACI-1548562, and in particular Expanse at the San Diego Supercomputer Center through allocation TG-DMR180071.
K.B. acknowledges the support of the U.S. Department of Energy, Office of Science, Office of Advanced Scientific Computing Research, Department of Energy Computational Science Graduate Fellowship under Award Number DE-SC0020347.

The authors wish to thank Sabyasachi Tiwari, Zhenband Dai, Nikolaus Kandolf, and Hitoshi Mori for their contributions to the EPW project that are not discussed in this manuscript;
John Cazes and Hang Liu at TACC for their support with the Characteristic Science Applications project, Paolo Giannozzi for his support with \texttt{Quantum ESPRESSO}; and Stefano Baroni for fruitful discussions. S. P. would also like to thank Jae-Mo Lihm for useful discussions. 

\end{acknowledgments}

\clearpage
\newpage

\clearpage
\newpage

\clearpage
\newpage

\clearpage
\newpage

\clearpage
\newpage

\clearpage
\newpage

\clearpage
\newpage

\clearpage
\newpage

\clearpage
\newpage

\clearpage
\newpage

\def\bibsection{}
\section*{References}
\bibliography{bibliography.bib}

\end{document}